\documentclass[aps,prd,twocolumn,floatfix]{revtex4}

\usepackage{graphicx}
\usepackage{dcolumn}
\usepackage{bm}

\begin{document}


\def\question#1{{{\marginpar{\small \sc #1}}}}
\def\qq{$ q\bar q $}
\def\ss{$ s\bar s $}
\def\nn{$n\bar{n}$}
\def\uu{$u\bar u$}
\def\dd{$d\bar d$}
\def\cc{$c\bar c$}
\def\cs{$c\bar s$}
\def\cq{$c\bar q$}
\def\cd{$c\bar d$}

\def\be{\begin{equation}}
\def\ee{\end{equation}}
\def\bd{\begin{displaymath}}
\def\ed{\end{displaymath}}
\def\ba{\begin{eqnarray}}
\def\ea{\end{eqnarray}}
\def\lr{\leftrightarrow}
\def\nn{$n\bar n$ }
\def\ccbar{$c\bar c$ }
\def\x{\bf x}
\def\B{\rm B}
\def\D{\rm D}
\def\E{\rm E}
\def\F{\rm F}
\def\G{\rm G}
\def\H{\rm H}
\def\I{\rm I}
\def\J{\rm J}
\def\K{\rm K}
\def\L{\rm L}
\def\M{\rm M}
\def\P{\rm P}
\def\S{\rm S}
\def\T{\rm T}
\def\V{\rm V}

\title{\Large Dynamics and Decay of Heavy-Light  Hadrons}
\author{ F.E. Close\footnote{\tt email:f.close@physics.ox.ac.uk}  and E.S. Swanson\footnote{On leave from the Department of Physics and Astronomy, University of Pittsburgh, Pittsburgh PA 15260. {\tt email: swansone@pitt.edu}} }
\affiliation{
Rudolph Peierls Centre for Theoretical Physics, Oxford University, Oxford,
OX1 3NP, UK.}

\date{\today}

\begin{abstract}

Recent signals for narrow hadrons containing heavy and light flavours are compared with
quark model predictions for spectroscopy, strong decays, and radiative transitions.
In particular, the production and identification of excited charmed
and \cs~ states
are examined with emphasis on elucidating the nature of $0^+$ and $1^+$ states. 
Roughly 200 strong decay amplitudes of $D$ and $D_s$ states up to 3.3 GeV are 
presented. Applications include  determining flavour content in $\eta$ mesons
and the mixing angle in $P$ and $D$ wave states and probes of putative molecular states.
We advocate searching for radially excited $D_s^*$ states in $B$ decays.

\end{abstract}  


\maketitle

\section{Introduction}

\begin{figure}[ht]
\includegraphics[scale=0.45,angle=0]{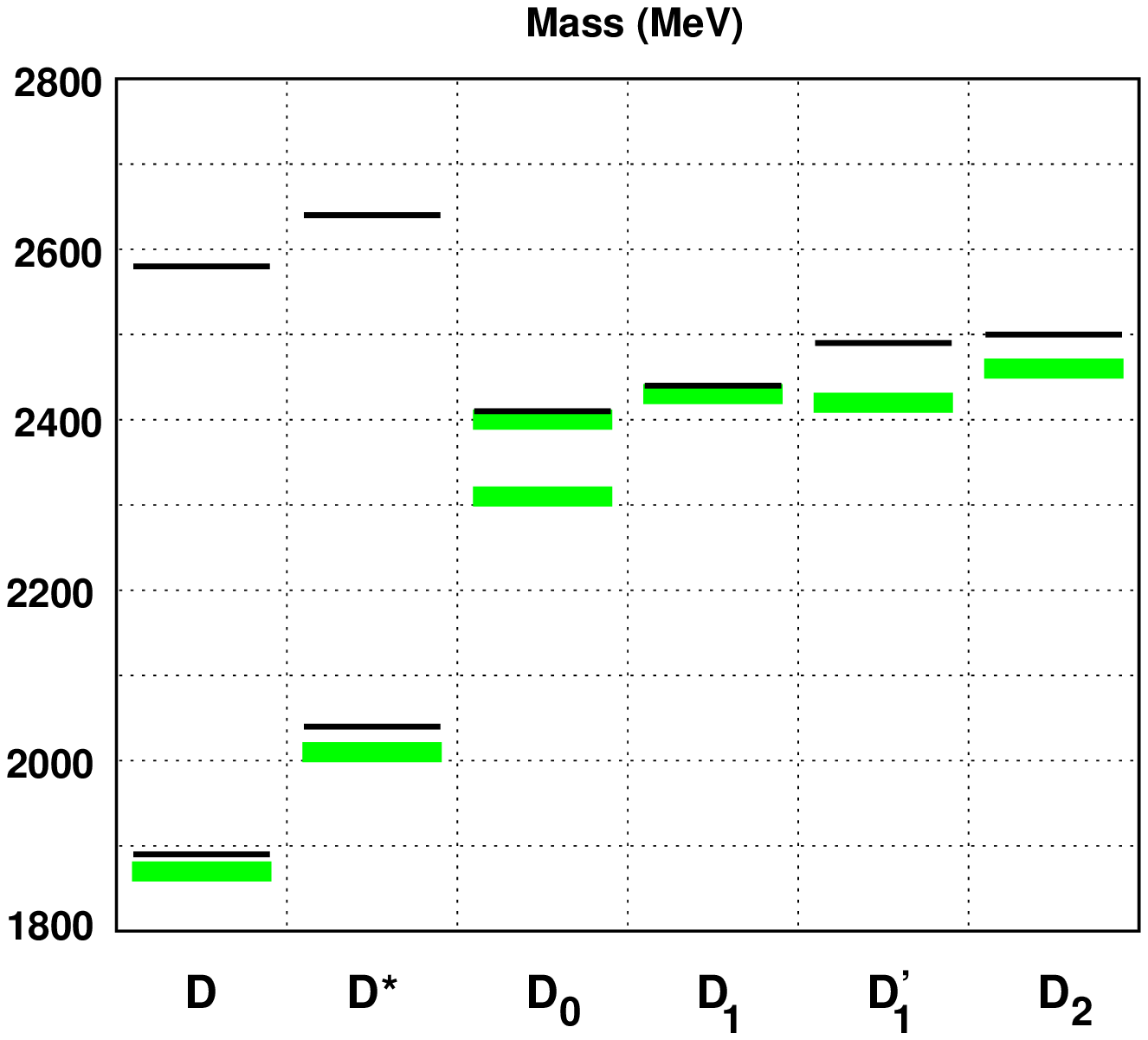}
\caption{\label{spect} $D$ Spectrum. The lines are
predictions from Ref. \protect\cite{GI}; the boxes are data \protect\cite{PDG}. Both the Belle and Focus $D_0$ states are shown \protect\cite{belle}.}
\end{figure}

\begin{figure}[ht]
\includegraphics[scale=0.45,angle=0]{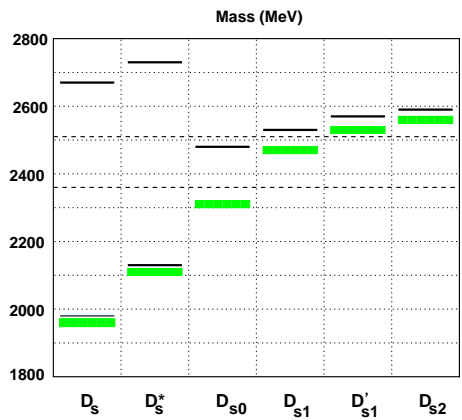}
\caption{\label{spect2} $D_s$ Spectrum. See Fig. \ref{spect}.The dashed lines are
the $DK$ and $D^*K$ thresholds.
}
\end{figure}

The discovery of the $D_s(2317)$ $(0^+)$ and $D_s(2460)$ $(1^+)$ mesons\cite{dsstates,PDG}, with masses considerably
lower than expected in potential models\cite{GI}, stimulated a range of 
theoretical activity. 
The current knowledge of charmed and \cs~ mesons, compared to the expectations of Ref \cite{GI} 
is summarised in Figs. \ref{spect},\ref{spect2} for the $D$ and $D_s$
systems respectively. There is a significant  amount of consistency  between theory 
and experiment,
interspersed with additional states that do not fit well
into such a classification, of which the $D_s(0^+)$ and $D_s(1^+)$ are particularly 
sharp examples. 
Attempts to accommodate these states have invoked a variety of mechanisms. One interpretation 
is that they
are indeed \qq $^3P_J$ levels, with  
their low masses being a realisation of chiral symmetry such that $m(0^+) - m(0^-) \equiv m(1^+) - m(1^-)$\cite{chiral}.
An alternative is that they are multiquark or molecular configurations
\cite{bcl} associated with the $DK$ and $D^*K$ thresholds. One unresolved issue in the latter 
class of models is whether
there are also further $(0,1)^+$ broad \cs~ states {\it above} $D^{(*)}K$ threshold, analogous to
what appears to occur with light flavoured scalar mesons\cite{CT02}.  
To help decide among competing interpretations, a coherent study of the dynamics of heavy-light hadrons is merited.

A particular issue in testing these hypotheses will be to determine the $^3P_1 - {}^1P_1$ mixing angle for the axial mesons.
In the chiral symmetry picture\cite{chiral} this is implicitly assumed to be the ideal heavy quark limit (see Section \ref{hqSect}).
In the molecular picture it is moot whether there is any simple mixing angle involving 
the $D_s(2460)$ and $D_s(2535)$
or whether a further axial with mass $\sim 2.5$GeV is called for.

Subsequent to the above discoveries, SELEX\cite{selex} reported a narrow state $D_s(2632)$ 
seen in $D_s \eta$ and $DK$ with a 
branching ratio $D_s(2632) \to D_s \eta \sim 6  DK$. The narrow width and the anomalous 
branching ratios (the two channels share the
same quark flavours and phase space favours the
$DK$ mode over the $D_s\eta$) led to
suggestions that this state may be a tetraquark\cite{maiani}. Within the more conservative 
\cs~ picture it was noted that the 
radially excited $2^3S_1$ is predicted
to lie at $\sim 2.73$ GeV
and that the presence of nodes in the wavefunction could lead to
suppression of certain modes if the decay momentum coincides with a node in 
momentum space\cite{bcdgs}. Such dynamics have been applied to the
decays of excited \cc~ states\cite{excitedcc} and also to light flavours with some 
success\cite{Barnes:1996ff}; however, Ref.\cite{bcdgs} found
that such an explanation would only work if extreme values for the parameters were 
chosen and thereby concluded that the SELEX state
might be an artefact. While we still agree with that conclusion, it does raise the 
possibility that narrow states could in principle occur if their masses and decay 
kinematics 
cause the momenta to coincide with
nodes; this is one of the questions
that we pursue in this survey. 

Independent of these tantalising possibilities, the transitions among excited hadrons 
can determine their dynamics and discriminate among models. With the production of charm 
in B decays, and the possible advent of charm factories at CLEO-c and GSI,
it is timely to assess the landscape for heavy-light hadrons.

Some highlights of the results are as follows:

Significant production of $D_s(2^3S_1)$ and $D_s(3^3S_1)$ is predicted in B decays.
Similarly, B decays may be a significant source of $D_1$ mesons and can be used
to test axial mixing.

Decay ratios such as ${D_s^*}' \to D_s \eta$/${D^*}' \to D\eta$ are useful probes
of the flavour structure of the $\eta$ meson.

The decays ${D_s^*}'' \to D_{s1}\eta$ and $D_{s1}'\eta$ test the putative molecular
nature of the $D_s(2460)$. Similarly, the decays $D_S'' \to D_{s0}\eta$, $D'' \to D_{s0}K$,
and ${D_s^*}'' \to D_{s0} K^*$ probe the structure of the enigmatic $D_s(2317)$.

There may be considerable spectroscopic mixing between the $D_s(2^3S_1)$ and $D_s(^3D_1)$
states which can be tested by measuring the transitions to $DK$ and $DK^*$ from the
vector $D_s$ states.

E1 transitions such as $2^3S_1 \to {}^1P_{0,1}$ are useful probes of the $D_1$ and
$D_{s1}$ mixing angles.

Novel radiative transition selection rules are obtained in the heavy quark limit.

Radiative transitions such as $1^+ \to 0^+\gamma$, $0^+ \to 1^- \gamma$, 
$1^+ \to 0^-\gamma$ can test molecular models in the $c\bar s$ sector. 
For example, the width of the
intermolecular transition $1^+ \to 0^+\gamma \sim 17$  keV contrasts with the 
${\cal O}(1)$ keV rate predicted for \cs~ states.

Anomalous branching ratios of excited states, for example, $D_s'' \to D_2 K$ being
much larger than $D^*K$ or $DK^*$, may be used to probe the nodal structure of 
hadron wavefunctions.


We proceed with a description of the strong decay model and the conventions concerning
state mixing. This is followed by a discusson of the phenomenology of the strong decay 
results and their relationship to the heavy quark limit. The penultimate section concerns
radiative transitions and highlights their utility as a diagnostic tool. All transition
rates are contained in Appendices B and C; Tables \ref{E1D}, \ref{E1Ds}, and
\ref{M1D} for radiative transitions and Tables \ref{DecTab1}--\ref{DecTab8} 
for strong transitions.

\section{Method}

\subsection{The Decay Model}

The $^3P_0$ model of strong decays assumes that $q\bar q$ pairs are created with
vacuum quantum numbers\cite{Micu:1968mk}. Thus the
interaction may be written as

\be
H_{q \bar q} = \gamma\, \sum_f 2 m_f \int d^{\, 3} x\; \bar \psi_f \psi_f \ ,
\ee
where $\psi_f$ is a Dirac quark field of flavour $f$, $m_f$ is the constituent quark mass, 
and $\gamma$ is a dimensionless \qq pair production strength.


Recent variants of the $^3P_0$ model consider modifications of the 
pair-production vertex~\cite{Roberts:1997kq}, or assume that pair creation
originates in a gluonic flux tube\cite{Kokoski:1985is}.
The latter is the ``flux-tube decay model", which in practice 
gives very similar predictions 
to the $^3P_0$ model.

The model has been extensively applied to meson and baryon 
strong decays, with considerable success\cite{applications,Barnes:1996ff,Godfrey:1986wj}.
The pair-production strength parameter $\gamma$ is fitted to strong decay data, 
and is roughly flavor-independent for decays involving production of
$u\bar u$, $d\bar d$ and $s\bar s$ pairs.  A typical value obtained from computation 
of light
meson decays is $\gamma = 0.4$\cite{Ackleh:1996yt,Barnes:1996ff,Barnes:2002mu},
assuming simple harmonic oscillator (SHO) wavefunctions with a
global scale, $\beta = 0.35$ -- $0.4$ GeV. The present work also assumes SHO wavefunctions 
but applies the formalism
to a variety of heavy-light mesons; thus we have allowed the SHO $\beta$ values to 
vary according to the state (see Table \ref{betaTab} in Appendix \ref{App1}). 
These values were obtained
by equating the root mean square (RMS) radius of the SHO wavefunction to that obtained in a simple
nonrelativistic quark model with Coulomb+linear and smeared hyperfine interactions.
Details are provided in the Appendix. 

In view of this it is appropriate to refit experimental data to obtain a new value for the
coupling, $\gamma$. A total of 32 experimentally well-determined decay rates have been 
fit with the model. Several variations of the decay model were also examined. Details
and the final parameters and method are discussed in Appendix \ref{App1}.

\subsection{Mixed States}
\label{Pwaves}

Heavy-light mesons are not charge conjugation eigenstates and so mixing can occur among 
states with the same $J^P$
that are forbidden for neutral states. These occur between states with $J = L$ and $S=1$ 
or $0$.  For example
the $J^{P} = 1^+$ axial vector $c\bar n$ and $c\bar s$ mesons
$\D_1$ and $\D_1'$ are coherent superpositions
of quark model $^3P_1$ and $^1P_1$ states,

\begin{eqnarray}
&|\D_1\rangle  &= +\cos(\phi) |^1\P_1\rangle + \sin(\phi) | ^3\P_1\rangle \nonumber \\
&|\D_1'\rangle  &= -\sin(\phi) |^1\P_1\rangle + \cos(\phi) | ^3\P_1\rangle.
\label{D1}
\end{eqnarray}
Quantifying the mixing pattern
as a function of flavour will give information about the internal dynamics; little is 
known empirically at present.
In the heavy quark limit $M_Q \to \infty$ there is an explicit prediction for
the mixing angle assuming that it is generated by spin-orbit interactions (see 
Section \ref{hqSect}).
One of our aims will be to devise tests for 
determining this mixing in practice for heavy, but finite flavour masses. 

Mixing between $S=0,1$ states with the same $J$ also occurs for
$^3D_2$ and $^1D_2$ states:
\begin{eqnarray}
|D_2^*\rangle &=& + \cos(\phi_D) |^1D_2\rangle + \sin(\phi_D) |^3D_2\rangle \nonumber \\
|{D_2^*}'\rangle &=& -\sin(\phi_D) |^1D_2\rangle + \cos(\phi_D) |^3D_2\rangle
\end{eqnarray}

Kokoski and Godfrey\cite{Godfrey:1986wj} find the angles (these supersede those of Ref. \cite{GI}) $\phi(1P_{cu}) = -26$ degrees and $\phi(1P_{cs}) = -38$ degrees, upon converting their
mixing conventions to ours. $D$-wave mixing angles were not computed.

Finally, the physical $\eta$ and $\eta'$ are taken to be

\begin{eqnarray}
\eta &=& +\cos(\theta) {1\over \sqrt{2}}(u \bar u + d\bar d) + \sin(\theta)s\bar s \nonumber \\
\eta' &=& -\sin(\theta) {1\over \sqrt{2}}(u \bar u + d\bar d) + \cos(\theta)s\bar s.
\end{eqnarray}
A typical quark model $\eta$ mixing angle is $\theta = -45$ degrees.
See Section \ref{etaSect}  for further discussion
on how to interpret the amplitudes for $\eta$ production.

\section{Strong Decays}

We proceed with a discussion of OZI allowed strong decays of $D$ and $D_s$ states.

\subsection{1S States}

$D^*$ decays are interesting because the open channels are so close to threshold
that isospin symmetry breaking mass shifts become important. In particular $D^{*+}$
may decay to both $D^0\pi^+$ and $D^+\pi^0$, but the $D^\pm\pi^\mp$ mode of $D^{*0}$
is closed. 
The experimental $D^{*+}$ widths are in the ratio

\begin{equation} 
{\Gamma(D^{*+} \to D^0 \pi^+) \over\Gamma(D^{*+} \to D^+ \pi^0)} = 2.21 \pm 0.57.
\end{equation}
The final state relative momenta are $q = 0.0393$ GeV and $q= 0.0379$ GeV
respectively, thus the form factor is essentially unity, the ratio
is dominated by the isospin factor and is predicted to be 2.28.

Absolute rates are given in
Table \ref{DecTab1} where one sees that the $D^{*+}$ widths are under-predicted by a factor 
of three. This is the largest error we have encountered with the $^3P_0$ model;
for example the analogous decay $K^* \to K \pi$ is under-predicted by approximately
45\% in amplitude.
While this would be easy to correct by adjusting the $^3P_0$ coupling,
we choose to retain the fit value of Appendix \ref{App1} since it does well globally.
We note that the decay model is tuned for SHO wavefunctions and decays of momenta of
hundreds of MeV. Thus it is perhaps no surprise that the largest error seen in the $^3P_0$
model is seen in this extreme, near-threshold, decay.

\subsection{P-waves}
\label{hqSect}


It is convenient to discuss heavy-light mesons in the $jj$ coupling scheme. One
has $L_J = s_{1/2}$ for the heavy quark which must combine with the light quark
spin and angular momentum to form a total $J^P$ state. In the P-waves 
$\ell_j = P_{1/2}$ and $P_{3/2}$. Thus the
$j = 1/2$ states form a doublet with $J^P = (0,1)^+$ while the $j=3/2$ states form
a $J^P = (1,2)^+$ doublet.

The relationship of these states to those in the  $LS$ coupling scheme can be
determined once the heavy quark dynamics has been isolated and conventions have been
fixed. We chose to employ the
conventions of Ref.\cite{Barnes:2002mu}.
This reference also discusses the other conventions for the
mixing angle that have appeared in the literature.
In the heavy-quark limit a particular ``magic" mixing angle follows
from the quark mass dependence of the spin-orbit and
tensor terms, which is $\phi_{HQ} = -54.7^o$ $(35.3^o)$
if the expectation of the heavy-quark spin-orbit interaction
is positive (negative) \cite{Godfrey:1986wj}. Since the former implies that
the $2^+$ state is greater in mass than the $0^+$ state, and this agrees with 
experiment, we employ $\phi = -54.7^o$ in the following.
This implies

\begin{eqnarray}
| P_1 \rangle_{HQ} &=&  +{1\over\sqrt{3}}| ^1P_1\rangle - \sqrt{2\over 3}|^3P_1\rangle \nonumber \\
| P_1' \rangle_{HQ} &=&  +\sqrt{2\over 3}| ^1P_1\rangle + {1\over \sqrt{3}}|^3P_1\rangle.
\label{axmix}
\end{eqnarray}

In practice the empirical mixing for the $D$ and $D_s$ systems
is not yet known.
Quantifying this is one of the challenges that we discuss here for both \cs and \cq systems.
For example, our decay model makes specific predictions for certain amplitude
ratios:

\begin{eqnarray}
{\cal A}({}^1P_1 \to V Ps)_S &=& -{1\over \sqrt{2}} {\cal A}({}^3P_1 \to V Ps)_S \\
{\cal A}({}^1P_1 \to V Ps)_D &=& \sqrt{2} {\cal A}({}^3P_1 \to V Ps)_D \\
{\cal A}(2{}^3S_1 \to {}^1P_1 Ps)_S &=& -{1\over \sqrt{2}} {\cal A}(2{}^3S_1 \to {}^3P_1 Ps)_S \\
{\cal A}(2{}^3S_1 \to {}^1P_1 Ps)_D &=& \sqrt{2} {\cal A}(2{}^3S_1 \to {}^3P_1 Ps)_D \\
{\cal A}({}^3D_1 \to {}^1P_1 Ps)_S &=& \sqrt{2} {\cal A}({}^3D_1 \to {}^3P_1 Ps)_S
\label{ratios}
\end{eqnarray}
which underpin the eventual extraction of the mixing angles.
These relationships imply that the  heavy quark $P_1$ state of Eq. \ref{axmix} 
couples to $VPs$ in $S$-wave, whereas the $P_1'$ heavy quark state couples in 
$D$-wave.
Thus one expects the $D_1$ (${D_1}'$)  to be broad (narrow) in the  heavy quark limit.
Similarly, Eq. \ref{ratios} implies that the
the $D_1 \pi$ ($D_1' \pi$) mode will be large (small) in $2{}^3S_1$ decays 
and the $D_1 \pi$ ($D_1' \pi$) mode
will be small (large) in ${}^3D_1$ decays.

Table \ref{DecTab1} give the predicted widths of the $D_1$ and $D_1'$ states in terms of
the mixing angle ($c$ denotes $\cos\phi$). Equating these to the measured rates of
$329 \pm 84$ MeV and $19 \pm 5$ MeV gives very good fits for mixing angles of
$\phi \approx -55^o$ or  $35^o$. The  first angle is the solution
for a broad $D_1$ while the second is for a broad $D_1'$. Since these correspond
to $-54.7^o$ and $+35^o$ respectively, good agreement is obtained with the heavy
quark predictions, although distinguishing the two scenarios is impossible.
This agreement may be tested by measuring decays with $D_1$s in the final state. The 
decay tables indicate that the most promising such decays are $D(^3D_1) \to D_1\pi$ and 
$D_1'\pi$, ${D^*}'' \to D_1\pi$ and $D_1'\pi$, and ${D_s^*}'' \to D_1 K$ and $D_1' K$.


The situation for the $D_{s1}$ is less satisfactory because both the $D_{s1}$ and
$D_{s1}'$ states are expected to be narrow due to the limited phase space available
for the $D^*K$ channel. The heavy quark $^3P_0$ model prediction for the $D_{s1}'$ 
width is 800 keV. Model uncertainties can be removed by measuring the ratio 

\begin{equation}
{\Gamma(D_{s1}' \to D^*K)\over \Gamma(D_{s2} \to D^*K)} = 84 \cos^2\phi_s + 42 \sin^2\phi_s - 119\cos\phi_s \sin\phi_s.
\end{equation}
The $D_{s1}$ mixing angle may also be accessed through the decays ${D^*}'' \to D_{s1}K$ and
$D_{s1}'K$ and ${D_s^*}'' \to D_{s1} \eta$ and $D_{s1}'\eta$ although only the $D_{s1}K$ mode
presents a substantial branching fraction.

The $1^+$ \cs~ states can be directly produced by the $W$ axial current in 
$B \to \bar{D} D_{s1}$  or $\bar{D} D_{s1}'$\cite{od}, thus B factories offer the
possibility of studying these intriguing states. However, heavy quark
symmetry suppresses the $P_{3/2}$ $1^+$ decay constant\cite{ors} so that production
of the $D_{s1}'$ may be negligible.
Furthermore
there is no conserved vector current suppression of the scalar \cs~ 
due to the different masses of the $c$ and $\bar{s}$
in such transitions and hence the $D_{s0}$ may also be detectable.
The relative production of these states in $B$
decays will provide further insight into the relationship between the 
various \cs~ states with $J^P = 0^+,1^+$.
For example, 
the enigmatic $D_{s0}$ is produced in $D''$ and ${D^*}''$ decays to $D_{s0}K$ and
$D_{s0}K^*$ respectively; with the former having a branching fraction of approximately 17\%.
This is sufficiently large that it may worth looking for.
One expects that 
these branching fractions would be substantially lower if the $D_{s0}$ state were 
a molecule.

In addition to the curious $J^P = 0^+$ and $1^+$ states in the \cs~ system, there are also 
questions in the charmed $D$ states. A simple problem is the existence of incompatible
candidates for the $0^+$ states at 
2308 and 2407 MeV\cite{belle}. The \cd~ state may 
be produced directly in $B$ decay as above,
but with the additional penalty of Cabibbo suppression.  It is also possible to
produce the $D_0$ in  $D''$ and $D_s''$ decays.

Finally, 
as noted in the Introduction, it has been suggested that
the anomalously low masses of the $D_s(2317)$ and $D_s(2460)$  are consistent 
with breaking heavy quark and chiral symmetry\cite{chiral}. This implies
$m(0^+) - m(0^-) \equiv m(1^+) - m(1^-)$. Presumably this relationship applies
to $s_{1/2}$ and $p_{1/2}$ states and hence the mass of the broad $D_{(s)1}$ state should 
employed.  For the $D_s$ system one obtains 349 MeV and 347 MeV for the left and right sides
of the equation. Although not considered by the authors of Ref. \cite{chiral}, in principle 
this relationship applies to the $D$ system as well. The $J=1$ mass difference is measured
to be 347 MeV and the $J=0$ mass difference is 349 MeV if the lighter Belle $D_0$ mass
is used. Thus, if taken seriously, this relationship supports the Belle results.
We stress that it is important to test the applicability of this model through
measurements of other observables, including the $P_1$ mixing angle.


\subsection{2S and 1D waves}

All $J^P$ combinations of $D^*$ can be produced in decays such as $B \to D^* \ell \nu$, 
and the hadronic analogues,
though transitions to states where the
light degrees of freedom are in highly excited states will be suppressed by poor wavefunction
overlaps and restricted phase space. 
The production of excited $1^-$ and $0^-$ should be feasible
 as they can be emitted from the $W$ in weak transitions such as 
$B \to D_s^{(*)} \bar{D}$: the form
 factor that is relevant here is driven by the wavefunction at short distances. 
A simple quark model computation confirms that the wavefunction at the origin is not 
strongly suppressed as the principle quantum number increases, and the recoil
momentum is sufficiently low that there is very little dependence on the $B \to D$ 
transition form factor.
Thus, the rate $B\to D_s^*(n) \bar D$ depends primarily on the phase space and one
finds that the relative emission rates are  

\begin{equation}
B\to D_s^* \bar D : B\to {D_s^*}' \bar D : B\to {D_s^*}'' \bar D \approx 1 : 0.35 : 0.03
\end{equation}

As the branching ratio $b.r.(B\to D_s^* \bar D)$ and $b.r.(B\to D_s^* \bar {D}^*)$ are each 
approximately  1\%\cite{PDG} 
one expects a total production of $2^3S_1({D_s^*}')$ at roughly 1\%.
And even the doubly excited $D_s^*$ will have a branching fraction of order
$10^{-3}$.
The charmed analogue $2^3S_1({D^*}')$ is Cabbibo suppressed and can be expected 
at the $10^{-3}$ level.
As noted in the previous Section, the axial and scalar \cs~ states can be produced in this way. For example, the $0^-$ $D_s'$ can also be produced; its decay is predicted to be 
almost entirely into $D^*K$.
We therefore advocate that these states should be sought at high statistics $B$-factories.

The signatures for the vector states are as follows. The \cs~ $2^3S_1({D_s^*}')$ decays 
dominantly to 
$D^* K (80\%)$ and  $DK (15\%)$ with
 traces of $D_s \eta$ and $D_s^* \eta$. The $D_1K$ channel is closed; to access the $D_1$ 
this way requires the $3^3S_1$ initial state (see the discussion in the next section).
The charmed $2^3S_1({D^*}')$ decays dominantly to $D_1\pi (75\%)$ and $D^*\pi (10\%)$ with
traces of $D \pi$ and $D^*\eta$.


The first excited vector ${D_s^*}'$ and ${D^*}'$ arise in either $2^3S_1$ or $1^3D_1$ 
configurations and in general there can be mixing between these.
Such mixing will tend to shift the masses of the eigenstates away from the simple 
potential model values of Figs. \ref{spect},\ref{spect2}. The
unperturbed masses for $D_s^*$ of $2^3S_1(2.73)$ or $1^3D_1(2.90)$ imply that one of 
these eigenstates will be kinematically
forbidden to decay to $D(1P)K$ while the other will be allowed. In the latter case there 
are interesting nodal effects, whose
character will depend on the mixing angle.

Similar remarks hold for the \cq~ states and the decays to $D(1P)\pi$. However in this case the small pion mass implies that the
$D(1P)\pi$ channel may be open for both initial states.
The role of decays of these states in determining the $1^+$ mixing angles by decays to 
$1^+0^-$ was
discussed in Section \ref{hqSect}.

The $0^+$ 1P state is clearly $Qp_{1/2}$, so there is no mixing problem with this final state in the transition from
$2S (0^-) \to 0^+0^-$.
Thus the $0^+$ may be accessed via this transition from a $D_s(2S)$ produced in B decay.
  So given a $D_s(2S)$ emitted from the W current in B decay, one
may access the $0^+$ state by the above transition. However, the analogous $D'$ production is Cabibbo 
suppressed and it decays 
dominantly to $D^*\pi$; the $D_s'$ analogously
decays to $D^*K$. It would be interesting to
study transitions to the analogous $D_s$ states and determine whether the $D_s(2317)$ and 
$D_s(2460)$ are \cs~ or other compounds. This
would require the initial state to be 3S in order to be kinematically open.

Mixing in the $^3D_2 - {}^1D_2$ system may be addressed in the heavy quark 
limit of the constituent quark model.
Diagonalising the the spin-orbit interaction yields the results

\begin{eqnarray}
M(^3D_1) &=&  M(D_2^*) = M_0 - {3\over 2} \langle H_{SO}^q\rangle_D \\ \nonumber
M(^3D_3) &=&  M({D_2^*}') = M_0 + \langle H_{SO}^q\rangle_D ,
\end{eqnarray}
and a $^3D_2- {}^1D_2$ mixing angle of $\phi_D = -50.76^o$. 
Thus, the heavy quark $D$-wave states are

\begin{eqnarray}
| D_2^* \rangle_{HQ} &=&  \sqrt{2\over 5}| ^1D_2\rangle - \sqrt{3\over 5}|^3D_2\rangle \nonumber \\
| {D_2^*}' \rangle_{HQ} &=&  \sqrt{3\over 5}| ^1D_2\rangle + \sqrt{2\over 5}|^3D_2\rangle.
\label{D2mix}
\end{eqnarray}

As with $P$-waves, the $^3P_0$ strong decay model makes specific predictions for 
$D$-wave heavy
quark decay amplitudes which may be useful in interpreting the spectroscopy. Some of
these are:

\begin{eqnarray}
{\cal A}({}^1D_2 \to V Ps)_P &=& -\sqrt{2\over 3} {\cal A}({}^3D_2 \to VPs)_P \\
{\cal A}({}^1D_2 \to V Ps)_F &=& +\sqrt{3\over 2} {\cal A}({}^3D_2 \to VPs)_F \\
{\cal A}({}^1D_2 \to {}^1P_1 Ps)&=& 0
\end{eqnarray}
The second of these is an example of the $^3P_0$ selection rule forbidding such 
transitions among \qq spin singlets\cite{fcpage}. 

We note that, in analogy with the $P$-waves, the amplitude ratios above imply that the $D_2^*$
decays strongly in $P$ wave while the ${D_2^*}'$ decays only in $F$ wave and is thus
narrower than the $D_2^*$. As with $P$-waves, this conclusion agrees with spin
conservation in the heavy quark limit. Unfortunately, the ability to distinguish
the states is weakened by the many
other decay modes which exist for these states. 
Nevertheless, 
the $D_2^*$ and ${D_2^*}'$
have total widths which depend strongly on $\phi_D$ and measurement of any (or several)
of the larger decay modes will provide (over) constrained tests of the model and 
measurements
of the mixing angle.

Finally we note that the transitions 
 $^1D_2 \to VV$ and $^3D_2 \to VV$ proceed in $^3P_2$, $^5P_2$, $^3F_2$, and $^5F_2$ waves,
but never share a wave. Thus there is no $^1D-{}^3D$ mixing due to $VV$ loops.

\begin{figure}[h]
\includegraphics[scale=0.35,angle=-90]{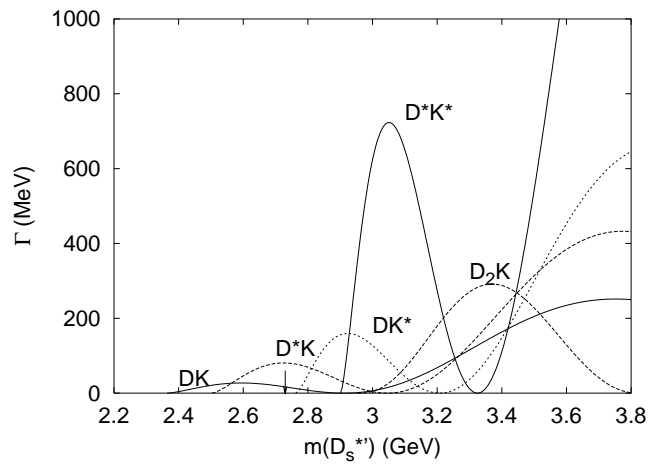}
\caption{\label{G_Ds_star_p} ${D_s^*}'$ Partial Widths vs. Mass. The arrow shows the
nominal mass of the ${D_s^*}'$.}
\end{figure}
\begin{figure}[h]
\includegraphics[scale=0.35,angle=-90]{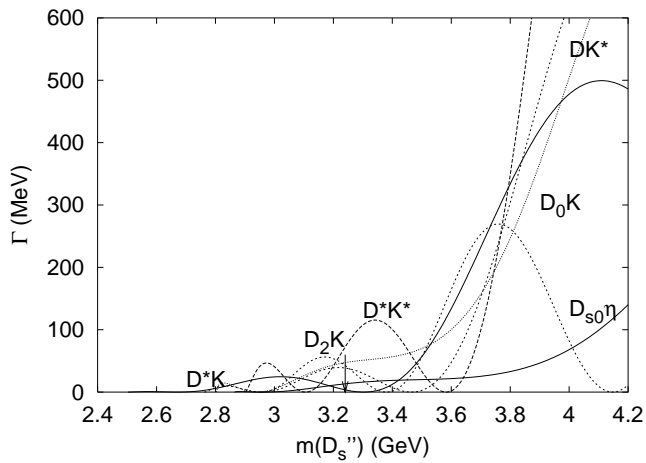}
\caption{\label{G_Dspp} ${D_s}''$ Partial Widths vs. Mass. The arrow shows the
nominal mass of the ${D_s}''$.}
\end{figure}

The effects of wavefunction nodes can be seen in Figs. \ref{G_Ds_star_p} and \ref{G_Dspp}. 
Here the
partial widths are plotted for fixed $\beta$ as a function of the mass of the
initial state.  For the ${D_s^*}'$, nodes would significantly affect the total width
if the mass were roughly 3.1 MeV. Alternatively, nodes directly affect the width 
of the $D_s''$ by suppressing the $D^*K$, $DK^*$, and $D^*K^*$ modes while
enhancing the $D_2K$ mode. The figure indicates that a $D_s''$ at 3.3 GeV would have
a substantial branching fraction to $D^*K^*$ while one at 3.4 GeV would have no
$D_2K$ mode. Clearly these effects must be accounted for in the phenomenology of 
heavy-light mesons.

\subsection{3S and 2D waves}

Similar remarks apply here as to the $2S$ waves with the bonus that phase space for decays into the $1P$ states is  open leading to potential measures of the axial mixing angle (Tables \ref{DecTab5}  and \ref{DecTab8}). The challenge is
to produce a significant sample of these $3S$ states in $B$-decays. Our estimate
of the branching ratio
$b.r.(B\to D_s^{*''} [\bar D + \bar D^{*+}]) \sim 10^{-3}$ suggests that this 
may be feasible.
Decays to $D_1 X$ are predicted to be $\sim 50\%$, which may provide a significant source of the axial charmed mesons and a measure of their mixing. 
Decays to $D_{s1}\eta$ and $D'_{1s}\eta$ may also be measured and used to test if the $D_s(2460)$
is the \cs~ partner of the $D_{s1}(2535)$ or an independent state such as a $D^*K$ molecule.

The decays of $D_s''$ and ${D_s^*}''$ can access the enigmatic $D_s(2317)$ and $D_s(2460)$. For example, comparison of the predicted rates for 
${D_s^*}'' \to D_{s1}\eta$ and ${D_s^*}'' \to D_{s1}'\eta$ 
will  test the putative \cs~ nature of the axial states.

In the charm $D$ system the decays  ${D_s^*}'' \to D_1 K$,  $D_1' K$ and $D_2 K$ 
measure the axial mixing angles. A robust prediction of the model is that the sum
of the two axial decay modes should be roughly $7.8$ times as large as the $D_2 K$
mode.

Finally, the status of $D_0(2308)$ and $D_0(2407)$ candidates may be tested by 
searching for them in the $D_0\pi$ and $D_0\eta$ decay modes of the $D''$ where they
have a non-negligible branching fraction.

\subsection{Probing the $\eta$ system}
\label{etaSect}

The emission of an $\eta$ in 
$2S \to 1S$ transitions is kinematically allowed, while $\eta'$ emission is forbidden.
Alternatively both are permitted in $3S \to 1S$ transitions. We note that the
flavour flow in the $D$ and $D_s$ systems probes the quark content of the $\eta$'s.
Thus $\eta$'s are produced via their \nn content in the \cq~system, whereas in \cs~
decays it is the \ss~components that are involved.

Thus a comparison of \cq~ $\to \eta$+\cq~ and \cq~$\to \pi+$\cq~probes the \nn~
content weighted by phase space and form factor effects. A 
rather direct measure of the \nn vs. \ss~content of the $\eta$ can be obtained 
by comparing
\[\Gamma\left( D^*(2S) \to D(1S) \eta(n\bar{n})\right ):\Gamma\left( D_s^*(2S) \to D_s(1S) \eta(s\bar{s})\right )
\]
or
\[
\Gamma\left( D(2S) \to D^*(1S) \eta(n\bar{n})\right ):\Gamma\left( D_s(2S) \to D_s^*(1S) \eta(s\bar{s})\right ).
\]
though the predicted branching ratios are small.


The predictions in Appendix C for $\eta$ and $\eta'$ production assume 
that these states 
are equally weighted mixtures
of \nn~ and \ss~. Hence
with conventions for the $\eta$ wavefunction specified above, the amplitude ratio $D_s \to D_s \eta$ and $D \to D \eta$ is
$\sqrt{2} \tan{\theta}$. This is modified by different wavefunctions, different quark
masses, and different meson masses. Direct computation gives the factors 
shown in Table \ref{etaTab}; these should be multiplied by $\sqrt{2}\tan{\theta}$ to
obtain physical amplitude ratios. Determining several of these amplitude ratios will
provide a constrained measure of the mixing angle and the efficacy of the decay model.

\begin{table}[h]
\caption{Amplitude ratios probing $\eta$ decays with $\sqrt{2}\tan{\theta}$ removed.}
\label{etaTab}
\begin{tabular}{llll}
\hline
${{D_s^*}' \to D_s \eta \over {D^*}' \to D \eta}$ &
${{D_s^*}' \to D_s^* \eta \over {D^*}' \to D^* \eta}$ &
${D_{s2}^{*} \to D_s \eta \over D_2^{*} \to D \eta}$ &
${D_s' \to D_s^* \eta \over D' \to D^* \eta}$ \\
\hline
1.78 & 0.918 & 1.09 & 0.483 \\
\hline
\end{tabular}
\end{table}

\section{Radiative Transitions}

Radiative transitions probe the internal charge structure of hadrons and are therefore 
useful in determining hadronic structure.
In particular they can help distinguish possible exotic molecular or
tetraquark state interpretations of the $D_s(0^+)$ and $D_s(1^+)$ and 
determine mixing angles.

\subsection{E1 and M1 Transitions}

E1 radiative partial widths are evaluated with the dipole formula
\begin{widetext}
\begin{equation}
\Gamma_{\rm E1}( nSLJ \to n'S'L'J' + \gamma)
 =  \frac{4}{3}\, C_{fi}\, \delta_{{\S}{\S}'} \, 
\left( { m_{\bar q} Q + m_q \bar Q \over m_q + m_{\bar q}}\right)^2 \,
\alpha \, |\,\langle nLJ| r|n'L'J' \rangle\, |^2 \, \omega^3 \, \frac{E_f}{M_i}
\label{E1Eq}
\end{equation}
\end{widetext}
where $Q$ and $\bar Q$ are the quark and antiquark charges in units of $|e|$,
$\alpha$ is the fine-structure constant,
$\omega$ is the final photon energy,
$E_f$ is the final  state's total energy,
$M_i$ is the initial state's mass,
and the angular matrix element $C_{fi}$ is

\begin{equation}
C_{fi}=\hbox{max}({\L},\; {\L}') (2{\J}' + 1)
\left\{ { {{\L}' \atop {\J}} {{\J}' \atop {\L}} {{\S} \atop 1}  } \right\}^2 .
\end{equation}

Wavefunctions were obtained from a simple nonrelativistic quark model which employs a
Coulomb+linear central potential with an additional smeared hyperfine interaction. Tensor
and spin-orbit terms are neglected. 
Results for E1 and M1 radiative transitions assuming $q\bar q$ structure are given in Appendix \ref{AppE1}.

Since to leading order E1 transitions are diagonal in spin, they select the $^3P_1$ component of the 
$1_L$ and $1_H$ in processes such as $D_1 \to D^*\gamma$. Thus measuring these
rates yields a direct estimate of the $P$-wave mixing angle. The most promising
processes are $D_1' \to D^*\gamma \approx 800 \cos^2\phi$ and $D_1' \to D\gamma \approx 1100 \sin^2\phi$, since both are large  and involve the narrow $D_1'$. Prospects for using 
E1 transitions to measure $\phi_s$ are less promising since the $D_s$ rates are all ${\cal O}(10)$ keV.

\subsection{Molecular Probes}

The peculiar properties of the $D_{sJ}(2317)$ raise the possibility that this is either a $c\bar s$ state
with substantial admixture of the $KD$ continuum\cite{shiftDs}, a tetraquark state\cite{tetra}, or a $DK$ molecule\cite{bcl}.
In the latter
case it is suspected that the same dynamics give rise to a $D^*K$ resonance which 
may be identified with the $D_s(2460)$. 

\begin{figure}[h]
\includegraphics[scale=0.6,angle=0]{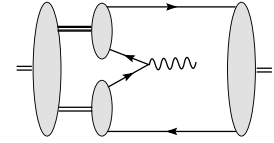}
\caption{\label{ANN} Molecular Radiative Transition.}
\end{figure}

The E1 radiative transitions $1^+ \to \gamma 0^-(c\bar{s})$ and 
$0^+ \to \gamma 1^-(c\bar{s})$
involve the overlap of molecular and \cs~ wavefunctions as shown in Fig. \ref{ANN}.
The amplitude is similar to one derived for the radiative decay of the 
$X(3872)$\cite{X2} and is given by

\begin{eqnarray}
{\cal A} &=& ({1\over \sqrt{2}}{2\over 3}e - {1\over \sqrt{2}}{1\over 3}e) \cdot \nonumber \\
&&\int d^3p\, d^3k\, \phi_{\rm mol}(p) \phi_D(k-{q\over 2} - \rho_{cu} {p\over 2}) \phi_K(k + {q\over 2} - \rho_{su} {p\over 2}) \cdot \nonumber \\
&& \phi_{D_s}^*(k+{p\over 2} + \rho_{cs} {q\over 2}) \cdot {\langle {\bf \sigma } \rangle \cdot \epsilon^*(q,\lambda)\over \sqrt{2q}} 
\end{eqnarray}
where
$\rho_{ij} = (m_i-m_j)/(m_i+m_j)$, $q$ is the momentum of the final state photon, and equal admixtures of the charged and
neutral components of the $J^P = 0^+$ or $J^P = 1^+$ $D^{(*)}K$ molecules has been assumed.
Evaluating this expression with SHO wavefunctions for the $D^{(*)}$, $K$, and $D_s$ 
states and setting the scale of the $D_s$ molecular wavefunction using the weak binding
relationship, $\langle r \rangle = 1/\sqrt{2 \mu_{DK}E_B} = \sqrt{3}/\sqrt{2}\beta_{WB}$ gives

\be
\Gamma(D_{s0}(mol) \to D_s^*\gamma) \approx 25 \ {\rm keV},
\ee
\be
\Gamma(D_{s1}(mol) \to D_s\gamma) \approx 30 \ {\rm keV},
\ee
and
\be
\Gamma(D_{s1}(mol) \to D_s^*\gamma) \approx 30 \ {\rm keV}.
\ee
These predictions  may be contrasted with the ${\cal O}(1)$ keV results 
for the analogous E1 transitions of simple \cs~ states reported in 
Table \ref{E1Ds}.

\begin{figure}[h]
\includegraphics[scale=0.8,angle=0]{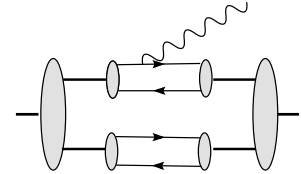}
\caption{\label{molmol} Molecule-Molecule Radiative Transition.}
\end{figure}

There is also the intriguing possibility of a radiative transition between the molecular
$D_s(D^*K)$ and $D_s(DK)$ states. In this case the rate is driven by a virtual $D^* \to D\gamma$ transition, as shown in Fig. \ref{molmol}.

The resulting rate is given by

\begin{eqnarray}
\Gamma_{MolMol} &=& {1\over 2}\Gamma(D^{*+} \to D^+\gamma;q_*) F_{+}(q_*) + \nonumber \\
               &&   {1\over 2}\Gamma(D^{*0} \to D^0\gamma;q_*) F_{0}(q_*) 
\end{eqnarray}
where the width for  $D^* \to D\gamma$ is evaluated at the photon momentum relevant to 
the $1^+ \to\gamma 0^+$ process, $q_*$
This result has been obtained assuming that the momentum space wavefunction for the 
molecular state is strongly peaked
at zero momentum, which will be the case for weakly bound states. The molecular form factors are  given by

\begin{equation}
F_\alpha(q) = \int {d^3k\over (2\pi)^3} \psi^{(\alpha)}_{1^+}(k) \psi^{(\alpha)*}_{0^+}(k - {m_K\over m_D+m_K}q)
\label{ff}
\end{equation}
where the wavefunctions refer to the bound $D^*K$ and $DK$ systems respectively and $\alpha$
is a channel index denoting the $D^{(*)+}K^0$ or $D^{(*)0}K^+$ components of the molecules.

Weak binding implies that $q_* \approx m_{D^*} - m_D \approx 140$ MeV.  
This is reduced by the mass factor $m_K/(m_D+m_K) = 0.2$
in the argument of the wavefunction in  Eq. \ref{ff}. Thus the product is
28 MeV which is
much smaller than the typical momenta in the weak binding potential, $\sqrt{2\mu_{DK}E_B} \approx 200$ MeV.
Thus the form factor
may be neglected and the predicted rate is given by the average of the neutral and charged
M1 transitions, $D^* \to D \gamma$. Consulting Table \ref{M1D} gives our final estimate
$\Gamma(1^+(mol) \to 0^+(mol) \gamma) \approx 17$ keV, which may give a measurable 
branching ratio. This may be contrasted with the
analogous $c\bar s$ transition predicted to be $(0.26\cos\phi_s + 1.4 \sin\phi_s)^2 \approx
1.0$ keV.

Measurement of the total widths of the $D_{s0}$ and $D_{s1}$ will be required before the
emerging data can be compared with these predictions\cite{Dexpts}.

\subsection{Radiative Transitions in the Heavy Quark Limit}

As there are no charge conjugation constraints on the $D$ and $D_s$ systems, radiative 
transitions from $^3S_1$ can reach both $^3P_J$ and $^1P_1$
states.  The most general transformation property of the transition at quark level is 
determined by noting that a positive helicity photon can change the projection of 
the spin or angular momentum of a quark by one unit.
For radiative transitions between $S$ and $P$ levels the most general transformation
for the current-quark operator is thus $AL_+ + BS_+ + CS_zL_+$, with unknown strengths $A,B,C$ (these can be calculated in specific models
but here we wish to make more general conclusions). The relative amplitudes for transitions to the various $J^P$ states are then
driven by Clebsch Gordan coefficients.
The procedure is defined in Refs.\cite{cot}
and gives for the relative radiative amplitudes to or from $(0,1,2)^+$ \footnote{In this section
the $D_1 $ and $D_{s1}$ ($D_1'$ and $D_{s1}'$) are referred to as $1_L$ ($1_H$)}:

\begin{eqnarray}
{\cal A}(V \to {}^3P_0 \gamma) &=& {1\over \sqrt{3}}\left(A-C - {B\over\sqrt{2}}\right) \\
{\cal A}(V \to 1_L\gamma)_{\lambda=0} &=&  {1\over \sqrt{2}}\left( (A-C)\sin\phi + B \cos\phi\right) \\
{\cal A}(V \to 1_L\gamma)_{\lambda=1} &=& \left(A-\frac{B}{\sqrt{2}}\right){\sin\phi\over \sqrt{2}} + C \cos\phi \\
{\cal A}(V \to {}^3P_2\gamma)_{\lambda=2} &=&  A+C\\
{\cal A}(V \to {}^3P_2\gamma)_{\lambda=1} &=& {1\over \sqrt{2}}\left(A+\frac{B}{\sqrt{2}}\right)\\
{\cal A}(V \to {}^3P_2\gamma)_{\lambda=0} &=& {1\over \sqrt{6}}(A-C) + {1\over \sqrt{3}}B
\end{eqnarray}


\noindent
Similar results hold for the decays involving the $1_H$.

These results simplify in the heavy quark limit, revealing some intriguing selection
rules. In particular, in the heavy quark limit one has $\phi = -54.7^o$, which implies

\begin{eqnarray}
-{\cal A}(V \to {}^3P_0\gamma) &=& {\cal A}(V \to 1_L\gamma)_{\lambda=0} \nonumber \\
                              &=& {\cal A}(V \to 1_L\gamma)_{\lambda =1} \\
{\cal A}(V \to 1_H \gamma)_{\lambda=0} &=& {\cal A}(V \to {}^3P_2\gamma)_{\lambda=0}  \\
{\cal A}(V \to 1_H \gamma)_{\lambda = 1} &=& \sqrt{2\over 3} {\cal A}(V \to {}^3P_2\gamma)_{\lambda=2} - \nonumber \\
&& \sqrt{1\over 3}{\cal A}(V \to {}^3P_2\gamma)_{\lambda=1} 
\label{HQsel}
\end{eqnarray}


The first of these corresponds to a selection rule which 
implies that the transition is pure $E1$ with $M2 \equiv 0$ (this is
immediately clear because the process $^3S_1 \to {}^3P_0\gamma$ has no M2 amplitude).
Hence
the equality of amplitudes implies this
is true for the $1_L$ rate as well.
These relationships also imply that (apart from phase space corrections)

\be
\Gamma(2^3S_1 \to 1_L \gamma ) = 3 \, \Gamma(2^3S_1 \to {}^3P_0\gamma)
\ee

\noindent 
and 

\be
\Gamma(1_L \to {}^3S_1\gamma) = \Gamma({}^3P_0 \to {}^3S_1\gamma).
\ee
 
\noindent 
Any deviation from these selection rules will measure deviations from the
heavy quark limit and the heavy quark mixing angle.

The selection rules can be understood more immediately in the heavy quark $jj$ basis. 
Only the light flavoured constituent can change its quantum state in the limit
$m_Q \to \infty$. In this case, the $^3S_1$ may be represented as 
$Q_{s(1/2)}\bar{q}_{s(1/2)}$ and the three independent $P$-states
are combinations of $Q_{s(1/2)}$ and $\bar{q}$ in $p_{1/2}$ and $p_{3/2}$ with $j_z = \pm 1/2$ or $p_{3/2}$ with $j_z = \pm 3/2$.
The three independent combinations of amplitudes in Eqs. \ref{HQsel} then correspond 
to the following:
the transition $s_{1/2} \to p_{1/2}$ controls the identical amplitudes to the $0^+$ and $1^+(s_{1/2}p_{1/2})$ states;
the transition $s_{1/2} \to p_{3/2}$ with $j_z = 1/2$ determines the second relation;
and 
the transition $s_{1/2} \to p_{3/2}$ with $j_z = 3/2$ controls the third relationship.

With a large enough sample of $2^3S_1$ $c\bar{s}$ states, the radiative amplitudes  to the $c\bar{s}$ $2^+$ and $1^+_H$ states
can be compared with the selection rules to determine the mixing angle for the $1_H$ state in the $^3P_1 - ^1P_1$ basis.
If the $1^+(2460)$ and $0^+(2317)$ are the remaining states in the $c\bar{s}$ 
$P$-wave system, the same mixing angle should emerge
when extracted from radiative transitions involving this pair of states. If $O(10^3)$ events are required to
measure radiative amplitudes, and each of these states is produced with $b.r.(\sim 10^{-3})$, then approximately $10^6$ initial $2^3S_1$ mesons are
required. Our estimates are that these arise at $\sim 10^{-2}$ in $B$ decays and so a suitable statistical sample should be 
accumulated at LHCb and other B-factories.

\section{Conclusions}

Excited $D$ and $D_s$ states beyond the $P$-wave have not yet been identified. Moreover,
mixing angles within the $P$ and $D$-waves are not yet quantified. Determining these 
observables is an important task for spectroscopy and heavy quark physics, and forms
a vital prerequisite for electroweak and CP violation studies.

We expect that the emission of radially excited $D_s$ and $D_s^*$ will be significant 
in $B$-decays and can be anticipated at the 1\% branching fraction level.
Their decays give a potential source of $P$-wave \cs~ states by which the $^3P-{}^1P$
mixing may be measured. This in turn can determine the $p_{3/2}-p_{1/2}$  mixing pattern,
which is important in testing models of the enigmatic $D_s(2317)$ and $D_s(2460)$ states.
For example 
chiral models implicitly assume that these are pure $p_{1/2}$ configurations while there
need be no simple mixing pattern in 
$D^{(*)}K$ molecular interpretations.
The $0^+$ and $1^+(p_{1/2})$ states will also be emitted in $B$-decays via the $W$ current. Their
relative rates and the presence of one or two axial mesons in the data can also test the
$p_{1/2}$ content and nature of the $D_s(2317)$ and $D_s(2460)$ states. Finally, radiative
transitions also probe the quark structure of these hadrons and can assist in distinguishing molecular and \cs~assignments.

Two body decay modes of excited $D_s^*$ states have nodes in simple nonrelativistic models.
Establishing the reality of these is important as such nodes have been invoked 
to explain anomalies in the \cc~ spectrum above charm threshold. The presence of nodes 
could in principle cause states to have narrow widths, though we find this unlikely in the
decays considered here unless 
unfavoured values of parameters are employed. In practice the nodes for certain decays 
tend to 
occur at energies where other channels have opened, thereby restoring a canonical width for the
state. However, the nodes are still implicitly manifested by virtue of the ensuing
anomalous branching ratios; for example, nodes might suppress the $DK$ channel while allowing a
$D^{(**)}K^{(*)}$ decay, leading to relative rates that are an inversion of ``phase space"
expectations. Thus
careful
measurement of the relative branching ratios for $3^3S_1$ $D_s$ decays could prove to be a 
powerful tool for understanding gluodynamics in quark models.

\acknowledgments

This work is supported, in part, by grants from the Particle Physics
and Astronomy Research Council, and the EU-TMR program ``Euridice''
HPRN-CT-2002-00311 (FEC) and by PPARC grant PP/B500607
and the U.S. Department of Energy under contract DE-FG02-00ER41135 (ESS).

\appendix

\section{Decay Computation Details}
\label{App1}

\subsection{Masses and SHO $\beta$ Values}

The evaluation of the perturbative decay amplitudes requires mesonic wavefunctions.
We follow tradition and employ SHO wavefunctions. Indeed, the model and experiment are 
sufficiently imprecise
that computations with more realistic quark model reveal no systematic 
improvements\cite{BG}.  The SHO wavefunction scale, denoted $\beta$ in the following,
is typically taken as a parameter of the model. However, since we seek to describe
the decay of heavy quark states, it is preferable to fix the SHO scales to quark
model wavefunctions. This was achieved by choosing $\beta$ to reproduce the RMS
radius of the quark model states. The resulting values are listed in Table \ref{betaTab}.

\begin{table}[h]
\caption{RMS Equivalent $\beta$ Values (GeV).}
\label{betaTab}
\begin{tabular}{c|llllll}
\hline
$n\,^{(2S+1)}L_J$ &  $uu$      &  $us$        & $ss$  &  $uc$  & $sc$ & $cc$ \\
\hline
$0\,^1S_0$      &  0.47 [0.4]  &  0.46 [0.4]  &  0.48 & 0.43 & 0.52 & 0.71   \\
$0\,^3S_1$      &  0.28        &  0.32        &  0.36 & 0.37 & 0.45 & 0.66   \\
$0\,^3P_J$      &  0.26        &  0.29        &  0.32 & 0.32 & 0.37 & 0.49   \\
$0\,^1P_1$      &  0.27        &  0.29        &  0.33 & 0.33 & 0.38 & 0.50  \\
$0\,^3D_J$      &  0.25        &  0.27        & 0.25  & 0.30 & 0.34 & 0.45  \\
$0\,^1D_2$      &  0.25        &  0.27        & 0.25  & 0.30 & 0.35 & 0.45   \\
$1\,^1S_0$      &  0.28        &  0.29        & 0.33  & 0.31 & 0.36 & 0.48   \\
$1\,^3S_1$      &  0.24        &  0.26        & 0.30  & 0.30 & 0.35 & 0.47  \\
$2\,^1S_0$      &  0.24        &  0.25        & 0.28  & 0.28 & 0.32 & 0.41   \\
$2\,^3S_1$      &  0.23        &  0.24        & 0.27  & 0.27 & 0.31 & 0.41   \\
\hline
\end{tabular}
\end{table}

The meson masses used to determine phase space and final state momenta are listed below.

Light meson masses:

$\pi = 0.138$, $\eta = 0.5477$, $\rho = 0.7758$, $\omega = 0.7826$, $K = 0.495$, $K^* = 0.8931$, $\eta' = 0.95778$.

$D$ meson masses:

$D = 1.8694$, $D^* = 2.0078$, $D_1 = 2.444$, $D_1' = 2.422$, $D_2 = 2.459$,
$D_0 = 2.308$ (Belle), $D_0 = 2.407$ (Focus).

Two experimental values for the scalar $D$ meson mass are reported \cite{belle}. Since
these are incompatible we prefer to compute with both masses; leaving it to
future experiment to choose between the options.

$D_s$ meson masses:

$D_s = 1.9683$,
$D_s^* = 2.1121$, $D_{s0} = 2.317$, $D_{s1} = 2.459$, $D_{s1}' = 2.535$, $D_{s2} = 2.572$.

Theoretical masses were: 

$D' = 2.58$, ${D^*}' = 2.64$, $D'' = 3.25$, ${D^*}'' = 3.31$, $D(^1D_2) = 2.83$, $D(^3D_1) = 2.82$, $D(^3D_2) = 2.83$,
$D(^3D_3) = 2.83$, $D_s' = 2.67$, ${D_s^*}'= 2.73$, $D_s'' = 3.24$, ${D_s^*}'' = 3.29$, 
$D_s(^1D_2) = 2.92$, $D_s(^3D_1) = 2.90$, $D_s(^3D_2) = 2.92$,
$D_s(^3D_3) = 2.92$.

These were obtained from the quark model used to determine the SHO scale or from 
Ref. \cite{GI}.

We set $D_1 = D_1(2444)$ and $D_1' = D_1(2422)$ since the latter is much narrower than the former. We also set
$D_{s1} = D_{s1}(2459)$ and $D_{s1}' = D_{s1}(2535)$ since both are narrow and the higher mass state is identified
with the $D_{s1}'$ in the heavy quark limit.

Finally, the quark model employed to determine the RMS $\beta$ values and the radiative
transition rates is a 
standard color Coulomb plus linear scalar confinement interaction with the addition
of 
a Gaussian-smeared contact hyperfine term.
The central potential is thus
\be
V(r) = \frac{4}{3}C  -\frac{4}{3}\frac{\alpha_s}{r} + br
+ 
\frac{32\pi\alpha_s}{9 m_1 m_2}\,
\tilde \delta_{\sigma}(r)\,
\vec {\S}_1 \cdot \vec {\S}_{2}\,
\ee  
where
$\tilde \delta_{\sigma}(r) = (\sigma/\sqrt{\pi})^3\, e^{-\sigma^2 r^2}$. The parameters
were chosen to reproduce a broad range of open flavour masses and are
$C_{uc} = -346$ MeV, 
$C_{sc} = -319$ MeV,
$b = 0.162$ GeV$^2$,
$\alpha = 0.594$, and
$\sigma = 897$ MeV.
Quark masses were taken to be $m_u = 0.33$ GeV, $m_s = 0.55$ GeV, and $m_c = 1.6$ GeV
in both radiative and strong computations.

\subsection{Parameter Determination}

A variety of $^3P_0$ models exist. These typically differ in the choice of weighting
function used in the pair creation vertex, meson wavefunctions employed, and the phase space conventions. We
shall restrict attention to the simplest vertex, which assumes a spatially uniform quark
creation probability density. Possible phase space conventions include relativistic
phase space (unit norm is used):

\begin{equation}
  {\rm (ps)} = 2\pi k {E_B E_C \over m_A} 
\end{equation}

\noindent
where $E_B$ is the energy of meson $B$ in the final state. This can differ substantially
from the nonrelativistic version:

\begin{equation}
  {\rm (ps)} = 2\pi k {m_B m_C \over (m_B + m_C)} 
\end{equation}

\noindent
especially when pions are in the final state. A third possibility, called the `mock meson'
method, is 
employed by Kokoski and Isgur\cite{Kokoski:1985is}:

\begin{equation}
  {\rm (ps)} = 2\pi k {M_B M_C \over M_A} 
\end{equation}

\noindent
where $M_A$ refers to the `mock meson' mass of a state. This is defined to be the
hyperfine-splitting averaged meson mass. In practice, the numerical result is little
different from the relativistic phase space except for the case of the pion, where a
mock mass of $M_\pi = 0.77$ GeV is used.  The final possibility is referred to as 'RPA
phase space'\cite{PSS} and postulates that the backward moving Fock components of 
pseudo-Goldstone bosons (pions and kaons) contribute to decays. In the chiral limit 
the net effect
of this is to multiply amplitudes containing a single pion or kaon by a factor of 2; if
two Goldstone bosons are present, the amplitude is multiplied by 3.

We have investigated the efficacy of six models in describing 32 well established
experimental decay widths. These models  all use SHO wavefunctions, either using
a universal SHO width of $\beta = 400$ MeV, the  RMS-equivalent $\beta$ values of Table \ref{betaTab},
or the RMS $\beta$s with the exception of $\beta_\pi$ and $\beta_K$ which are set to
400 MeV. The latter choice is an attempt to recognise that the lighter pseudoscalar
states are Goldstone bosons and hence are likely to be larger than simple quark
mode estimates. Relativistic and RPA phase space conventions have also been tested. We
remark that the RPA and mock meson prescriptions yield similar results.

The resulting best fit $^3P_0$ couplings and their errors are listed in Table \ref{couplings}. As can be seen, the data and model are of sufficiently low quality that it is very
difficult to distinguish the models (the models with $\beta_\pi = 0.4$ obtain 
$\delta \gamma/\gamma \approx 38\%$ while employing $\beta_{RMS}$ yields $\delta \gamma/\gamma \approx 29\%$).
We henceforth adopt the relativistic
phase space convention with  RMS $\beta$ values determined as in Table \ref{betaTab} and set
$\gamma = 0.485$.

\begin{table}
\caption{$^3P_0$ Couplings.}
\label{couplings}
\begin{tabular}{l|ccc}
\hline
phase space & $\beta_{RMS}$ & $\beta_\pi = 0.4$ & all $\beta = 0.4$ \\
\hline
Rel.        & $0.485 \pm 0.15$  &  $0.417 \pm 0.16$   &  $0.505 \pm 0.18$  \\
RPA         & $0.214 \pm 0.06$ &  $0.186 \pm 0.07$   &  $0.228 \pm 0.09$  \\
\hline
\end{tabular}
\end{table}

\begin{figure}[h]
\includegraphics[scale=0.35,angle=-90]{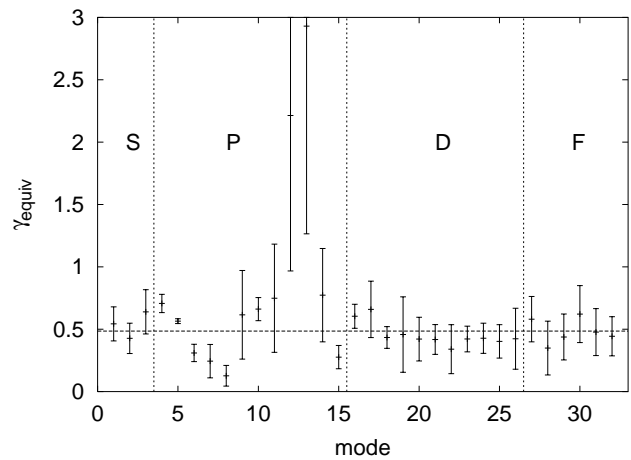}
\caption{\label{fits2} Equivalent Coupling vs. Mode. $^3P_0$ couplings required to reproduce
experiment for 32 decay modes.}
\end{figure}

The couplings required to reproduce experiment in the 32 decay modes for the model
used here are shown in Fig. \ref{fits2}. As can be seen, the large experimental errors
preclude definitive conclusions. Nevertheless the model provides clear guidance over 
three orders of magnitude of predicted widths and over a broad range of
quark flavours and meson quantum numbers. The specific decay modes are listed in Ref. \cite{modes}

\section{E1 and M1 Radiative Transitions}
\label{AppE1}

Radiative transition rates based on Eq. \ref{E1Eq} and the mixed states of 
Section \ref{Pwaves}
are reported here.

\begin{table}
\caption{E1 Radiative Transitions in the $D$ System}
\label{E1D}
\begin{tabular}{ccc}
\hline
mode & q (MeV) & $\Gamma$ (keV) \\
\hline
$D_2^+ \to D^{*+} \gamma$ & 408 & 51 \\
$D_2^0 \to D^{*0} \gamma$ & 410 & 895 \\
$D_0^+ \to D^{*+} \gamma$ & $\pmatrix{279 \\ 364}$ & $\pmatrix{17 \\ 37}$ \\
$D_0^0 \to D^{*0} \gamma$ & $\pmatrix{281 \\ 367}$ & $\pmatrix{304 \\ 649}$ \\
${D^*}{'^+} \to D_0^+\gamma$ & $\pmatrix{ 311 \\ 223}$  & $\pmatrix{10.0 \\ 3.8 }$ \\
${D^*}{'^0} \to D_0^0\gamma$ & $\pmatrix{ 311 \\ 223}$  & $\pmatrix{173 \\ 66 }$ \\
${D^*}{'^+} \to D_2^+\gamma$ & 175  & 9.4 \\
${D^*}{'^0} \to D_2^0\gamma$ & 175  & 163 \\
$D_1^+\to D^{*+}\gamma$ & 377  & $s^2$ 41 \\
$D_1^0\to D^{*0}\gamma$ & 379  & $s^2$ 715 \\
${D_1'}^+ \to D^{*+}\gamma$ & 395  & $c^2$ 46 \\
${D_1'}^0 \to D^{*0}\gamma$ & 398  & $c^2$ 819 \\
${D^*}'^+ \to D_1^+\gamma$ & 209  & $s^2$ 9.5 \\
${D^*}'^0 \to D_1^0\gamma$ & 209  & $s^2$ 164 \\
${D^*}'^+ \to {D_1}'^+\gamma$ & 189  & $c^2$ 7.0 \\
${D^*}'^0 \to {D_1}'^0\gamma$ & 189  & $c^2$ 122 \\
$D_1^+ \to D^+\gamma$ & 490  & $c^2$ 59 \\
$D_1^0 \to D^0\gamma$ & 493  & $c^2$ 1046 \\
${D_1}'^+ \to D^+\gamma$ & 507  & $s^2$ 66 \\
${D_1}'^0 \to D^0\gamma$ & 510  & $s^2$ 1154 \\
${D}'^+ \to D_1^+\gamma$ & 153  & $c^2$ 14 \\
${D}'^0 \to D_1^0\gamma$ & 153  & $c^2$ 233 \\
${D}'^+ \to {D_1}'^+\gamma$ & 132  & $s^2$ 8.8 \\
${D}'^0 \to {D_1}'^0\gamma$ & 132  & $s^2$ 152 \\

\hline
\end{tabular}
\end{table}

\begin{table}
\caption{E1 Radiative Transitions in the $D_s$ System}
\label{E1Ds}
\begin{tabular}{ccc}
\hline
mode & q (MeV) & $\Gamma$ (keV) \\
\hline
$D_{s2} \to D_s^{*} \gamma$ & 419 & 8.8 \\
$D_{s0} \to D_s^{*} \gamma$ & 196 & 1.0 \\
${D_s^*}' \to D_{s0} \gamma$ & 382 & 3.3\\
${D_s^*}' \to D_{s2} \gamma$ & 153 & 1.2\\
$D_{s1} \to D_{s}^* \gamma$ & 323 & $s^2$ 4.2\\
${D_{s1}}' \to D_{s}^* \gamma$ & 388 & $c^2$ 7.1\\
$D_{s1} \to D_{s} \gamma$ & 442 & $c^2$ 7.3\\
${D_{s1}}' \to D_{s} \gamma$ & 504 & $s^2$ 10.6\\
${D_s^*}' \to D_{s1} \gamma$ & 258 & $s^2$ 3.2\\
${D_s^*}' \to D_{s1}' \gamma$ & 188 & $c^2$ 1.3\\
$D_s' \to D_{s1} \gamma$ & 203 & $c^2$ 5.4\\
$D_s' \to D_{s1}' \gamma$ & 132 & $s^2$ 1.5\\
\hline
\end{tabular}
\end{table}

\begin{table}
\caption{M1 Radiative Transitions}
\label{M1D}
\begin{tabular}{cccc}
\hline
mode & q (MeV) & $\Gamma$ (keV) & $\Gamma$ (keV) PDG \\
\hline
$D^{*+} \to D^+ \gamma$ & 136 & 1.8 & $1.5 \pm 0.5$ \\
$D^{*0} \to D^0 \gamma$ & 137 & 32 & $ < 800 $ \\
$D_1^{+} \to D_0^+ \gamma$ & $\pmatrix{ 132 \\ 37}$  & $\pmatrix{(0.7 c + 1.8s)^2  \\( 0.11 c + 0.28 s)^2} $ & \\
$D_1^{0} \to D_0^0 \gamma$ & $\pmatrix{ 132 \\ 37}$  & $\pmatrix{(3.2 c + 2.0s)^2 \\ (0.47 c+ 0.31s)^2 }$ & \\
${D_1^{+}}' \to D_0^+ \gamma$ & $\pmatrix{ 111 \\ 15}$  & $\pmatrix{(-0.56 s + 1.4c)^2 \\ (-0.031 s + 0.07 c)^2}$ &  \\
${D_1^{0}}' \to D_0^0 \gamma$ & $\pmatrix{ 111 \\ 15}$  & $\pmatrix{(-2.4 s + 1.6 c 0^2 \\ (-0.12 s + 0.08 c)^2}$ &  \\
\hline
$D_s^{*} \to D_s \gamma$ & 139 & 0.2 & \\
$D_{s1} \to D_{s0} \gamma$ & 139 & $(0.26 c + 1.4 s)^2$ &  \\
$D_{s1}' \to D_{s0} \gamma$ & 209 & $(-0.44 s + 2.5 c)^2$ & \\
\hline
\end{tabular}
\end{table}

\section{Tables of Open Flavour Strong Decay Modes}

Strong decay rates and amplitudes are collected in Tables \ref{DecTab1} to \ref{DecTab8}.
In the following $c$ and $s$ refer to $P$-wave mixing in the $D$ or $D_s$ sector
$c = \cos\phi$ or $\cos\phi_s$. Mixing angles in the $D$-waves are labelled $c_1$ or
$s_1$; thus $c_1 = \cos\phi_D$ or $\cos\phi_{Ds}$. Estimates of decay rates containing
mixing angles are given in terms of the theoretical heavy quark predictions.
Rates for modes containing the Belle and Focus $D_0$ are presented separately; when 
both may occur
the differing possible total widths are separated by a slash.

\begin{table*}
\caption{Open-flavor strong decays}
\label{DecTab1}
\vskip 0.3cm
\begin{tabular}{llll}
State &  Mode & $\Gamma_{thy}$\ \  [$\Gamma_{expt}$] (MeV) & Amps. (GeV$^{-1/2}$) \\
\hline
\hline
$\D^{*+}$ &  $\D^0 \pi^+$  & 25 keV [64(15) keV]  & $^1P_1 = -0.027$  \\
$\D^{*+}$ &  $\D^+ \pi^0$  & 11 keV [29(6.8) keV]  & $^1P_1 = -0.019$  \\
$\D^{*0}$ &  $\D^0 \pi^0$  & 16 keV [$<$  2.1 MeV]  & $^1P_1 = -0.021$  \\
\hline
$\D_0(2308)$ & $\D\pi$ & 316 [276(66)] & $^1S_0 = +0.635$ \\
\hline
$\D_0(2407)$ & $\D\pi$ & 283 [276(66)] & $^1S_0 = +0.504$ \\
\hline
$\D_1$  & $\D^*\pi$ &  $110 c^2 + 191 s^2 - 230 sc = 272$ [329(84)] & $^3S_1 = -0.338 c + 0.478 s$, $^3D_1 = -0.153 c - 0.108 s$ \\
\hline
$\D_1'$  & $\D^*\pi$ &  $191 c^2 + 106 s^2 + 240 sc = 22$ [19(5)] & $^3S_1 = +0.504 c + 0.357 s$, $^3D_1 = -0.099 c + 0.141 s$\\
\hline
$\D_2$  & $\D\pi$ &  35 & $^1D_2 = +0.164$\\
        & $\D^*\pi$ &  20 & $^3D_2 = -0.154$\\
        & $\D\eta$ &  0.08 & $^1D_2 = +0.013$\\
        & total: &  55 [24(5)]  & \\
\hline
\hline
\end{tabular}
\end{table*}

\begin{table*}
\caption{Open-flavor strong decays, continued}
\label{DecTab2}
\vskip 0.3cm
\begin{tabular}{llll}
State &  Mode & $\Gamma_{thy}$\ \  [$\Gamma_{expt}$] (MeV) & Amps. (GeV$^{-1/2}$) \\
\hline
\hline
$\D'$ &  $\D^* \pi$  & 27   & $^3P_0 = -0.145$  \\
      &  $\D_0(2308) \pi$  & 41  & $^1S_0 = +0.355$  \\
     &  $\D_0(2407) \pi$  & 18  & $^1S_0 = +0.416$  \\
      &  $\D^* \eta$  & 1  & $^3P_0 = -0.051$  \\
      & total: & 69/46 & \\
\hline
$\D^{*'}$ & $\D\pi$ & 1 & $^1P_1 = -0.025$ \\
          & $\D^*\pi$ &  5 & $^3P_1 = -0.059$ \\
          & $\D\eta$  &  0.4 & $^1P_1 = +0.016$ \\
          & $\D_s K$  &  0.1 & $^1P_1 = +0.010$ \\
          & $\D^* \eta$ & 2  & $^3P_1 = -0.055$ \\
          & $\D_1 \pi$ &  $11 c^2 + 22 s^2 - 31 sc = 33$  & $^3S_1 = -0.270 c + 0.381s$  \\
          &            &                              & $^3D_1 = -0.034c - 0.024s$  \\
          & $\D_1' \pi$ &  $ 27 c^2 + 14 s^2 + 38 sc = 1$ & $^3S_1 = +0.369 c + 0.261 s$ \\
          &             &                                      & $^3D_1 = -0.035 c + 0.049 s$ \\
          & $\D_2 \pi$ &  0.1 & $^5D_1 = +0.031$ \\
          & $\D_s^* K$  &  0.7 & $^3P_1 = -0.041$ \\
          & total: & 44  & \\
\hline
\hline
\end{tabular}
\end{table*}

\begin{table*}
\caption{Open-flavor strong decays, continued}
\label{DecTab3}
\vskip 0.3cm
\begin{tabular}{llll}
State &  Mode & $\Gamma_{thy}$\ \  [$\Gamma_{expt}$] (MeV) & Amps. (GeV$^{-1/2}$) \\
\hline
\hline
$\D(^3D_1)$ &  $\D \pi$  & 73  & $^1P_1 = -0.161$  \\
            &  $\D^* \pi$  & 45   & $^3P_1 = -0.141$  \\
            &  $\D \eta$  & 16   & $^1P_1 = -0.082$  \\
            &  $\D_s K$  & 55   & $^1P_1 = -0.164$  \\
            &  $\D^*\eta$   & 9   & $^3P_1 = -0.071$  \\
            &  $\D_1\pi$   & $131 c^2 + 64 s^2 + 183 sc = 0.2$  & $^3S_1 = +0.442c + 0.312 s$  \\
            &             &                                   & $^3D_1 = +0.106 c + 0.051 s$  \\
            &  $\D_1'\pi$   & $61 c^2 + 127 s^2 - 176 sc = 189$   & $^3S_1 = +0.288 c - 0.409 s$  \\
            &              &                                       & $^3D_1 = +0.057c - 0.115 s$  \\
            &  $\D_2\pi$   & 7   & $^5D_1 = +0.108$  \\
            &  $\D_s^* K$  & 23   & $^3P_1 = -0.129$  \\
            &  $\D \rho$  & 74  & $^3P_1 = -0.207$ \\
            &  $\D \omega$  & 16   & $^3P_1 = -0.098$   \\
            &  $\D^* \rho$  & 13   & $^1P_1 = -0.117$, $^3P_1 = 0$, $^5P_1 = 0.052$, $^5F_1 = 0.025$   \\
            &  $\D^* \omega$  & 3   & $^1P_1 = -0.063$, $^3P_1 = 0$, $^5P_1 = 0.028$, $^5F_1=  0.011$ \\
            & total: & 523  &  \\
\hline
$\D(^3D_3)$ &  $\D \pi$  & 53   & $^1F_3 = -0.136$  \\
            &  $\D^* \pi$  & 55   & $^3F_3 = +0.155$  \\
            &  $\D \eta$  & 4   & $^1F_3 = -0.041$  \\
            &  $\D_s K$  & 4   & $^1F_3 = -0.044$  \\
            &  $\D^*\eta$   & 3   & $^3F_3 = +0.037$  \\
            &  $\D_1\pi$   & $3.5 c^2 + 0.5 s^2 - 2.3 sc = 3$   & $^3D_3 = +0.071 c - 0.026s$  \\
            &              &                                      & $^3G_3 = +0.013 c + 0.009 s$  \\
            &  $\D_1'\pi$   & $0.6 c^2 + 4.6 s^2 + 2.8 sc = 2$    & $^3D_3 = -0.027 c - 0.077s$ \\
            &               &                                        & $^3G_3 = +0.011 c - 0.016s$ \\
            &  $\D_2\pi$   & 6   & $^5D_3 = -0.099$, $^5G_3 = -0.010$  \\
            &  $\D_s^* K$  & 2   & $^3F_3 = +0.034$  \\
            &  $\D \rho$  & 15   & $^3F_3 = +0.090$  \\
            &  $\D \omega$  & 4   & $^3F_3 = +0.050$  \\
            &  $\D^* \rho$  & 99   & $^5P_3 = -0.009$, $^1F_3 = 0.343$, $^3F_3 = 0$, $^5H_3 = 0.019$  \\
            &  $\D^* \omega$  & 27   & $^5P_3 = -0.004$, $^1F_3= 0.188$, $^3F_3 = 0$, $^5H_3 = 0.009$  \\
            &  $\D \eta'$  & $\approx 0$   & $^1F_3 = -5.5\cdot 10^{-5}$  \\
            & total: & 277 &  \\
\hline
\hline
\end{tabular}
\end{table*}

\begin{table*}
\caption{Open-flavor strong decays, continued}
\label{DecTab4}
\vskip 0.3cm
\begin{tabular}{llll}
State &  Mode & $\Gamma_{thy}$\ \  [$\Gamma_{expt}$] (MeV) & Amps. (GeV$^{-1/2}$) \\
\hline
\hline
${\D_2^*}$ &  $\D^* \pi$  & $93 c_1^2 + 91 s_1^2 +9 s_1c_1 = 87$   & $^3P_2 = +0.123 c_1 - 0.151 s_1$  \\
           &              &                                     & $^3F_2 = +0.158 c_1 + 0.129 s_1$  \\
            &  $\D_0(2308) \pi$  & $0.8 c_1^2 +  1.9 s_1^2 + 2.4 s_1c_1 = 0.4 $   & $^1D_2 = -0.026 c_1 - 0.041 s_1$  \\
            &  $\D_0(2407) \pi$  & $0.9 c_1^2 + 0.7 s_1^2 + 1.6 s_1c_1 = 0.02$   & $^1D_2 = -0.034 c_1 - 0.031 s_1$  \\
            &  $\D^*\eta$   & $13 c_1^2 + 18 s_1^2 - 22 s_1c_1 = 27$   & $^3P_2 = +0.077 c_1 - 0.094 s_1$  \\
            &              &                                    & $^3F_2 = +0.038 c_1 + 0.032 s_1$  \\
            &  $\D_1\pi$   & $0.7 c_1^2 s^2 + 5.1 s_1^2 c^2 + 7.6 s_1^2 s^2 + $   & $^3D_2 =-0.032 c_1 s - 0.087 s_1 c - 0.107 s_1s$  \\
            &              & $3.8 c_1s_1 s c + 4.6 c_1s_1 s^2 + 12.5 s_1^2 sc = 0.1$   &  \\
            &  $\D_1'\pi$  & $0.9 c_1^2c^2 + 9.8 s_1^2 c^2 + 6.8 s_1^2 s^2 + $   & $^3D_2 = -0.035 c_1c + 0.096 s_1 s - 0.115 s_1 c$  \\
            &              & $6.0 c_1s_1 c^2 - 5.0 c_1 s_1 s c -16.3 s_1^2 sc = 8.4$   &   \\
            &  $\D_2\pi$   & $86 c_1^2 + 115 s_1^2 - 195 s_1c_1 = 197$   & $^5S_2 = -0.349 c_1 + 0.427 s_1$  \\
            &              &                                    & $^5D_2 = -0.130 c_1 + 0.067 s_1$  \\
            &              &                                    & $^5G_2 = -0.013 c_1 - 0.011 s_1$  \\
            &  $\D_s^* K$  & $31 c_1^2 + 45 s_1^2 - 69 s_1c_1 = 73$   & $^3P_2 = +0.142 c_1 - 0.174 s_1$  \\
            &              &                                   & $^3F_2 = +0.035 c_1 + 0.029 s_1$  \\
            &  $\D \rho$  & $75 c_1^2 + 100 s_1^2 - 122 s_1c_1 = 150$   & $^3P_2 = +0.183 c_1 - 0.224 s_1$  \\
            &         &                                        & $^3F_2 = +0.092 c_1 + 0.075 s_1$  \\
            &  $\D \omega$  & $25 c_1^2 + 33 s_1^2 - 41 s_1c_1 = 50$  & $^3P_2 = +0.106 c_1 - 0.130s_1$  \\
            &         &                                        & $^3F_2 = +0.051 c_1 + 0.042 s_1$  \\
            &  $\D^* \rho$  & $50 c_1^2 + 25 s_1^2 = 33$   & $^3D_2 = -0.243 c_1$, $^5D_2 = 0.172 s_1$ \\
            &               &                          & $^3F_2 = -0.025 c_1$, $^5F_2 = 0.029 s_1$  \\
            &  $\D^* \omega$  & $14 c_1^2 + 7 s_1^2 = 9$   & $^3D_2 = -0.133 c_1$, $^5D_2 = 0.094 s_1$ \\
            &                 &                      & $^3F_2 = -0.012 c_1$, $^5F_2 = 0.014 s_1$  \\
            &  $\D_{s0} K$ &  $\approx 0$  &  $^1D_2 = -0.007 c_1 - 0.003 s_1$   \\
            & total: & $389 c_1^2 + 437 s_1^2 - 409 s_1c_1$ + 8 &  \\
\hline
${\D_2^*}'$ &  $\D^* \pi$  & $91 c_1^2 + 93 s_1^2 -9 s_1c_1 = 96$   & $^3D_2 = -0.151 c_1 - 0.123 s_1$  \\
           &               &                                      & $^3F_2 = +0.129 c_1 - 0.158 s_1$  \\
            &  $\D_0(2308) \pi$  & $1.9 c_1^2 + 0.8 s_1^2 - 2.4 s_1c_1 = 2.3$   & $^1D_2 = -0.041 c_1 + 0.026 s_1$  \\
            &  $\D_0(2407) \pi$  & $ 0.7 c_1^2 + 0.9 s_1^2 - 1.6 s_1c_1 = 1.6$   & $^1D_2 = -0.031 c_1 - 0.034 s_1$  \\
            &  $\D^*\eta$   & $18 c_1^2 + 13 s_1^2 + 22 s_1c_1 = 4.5$   & $^3D_2 = -0.094 c_1 - 0.077 s_1$  \\
            &             &                                    & $^3F_2 = +0.031 c_1 - 0.038 s_1$  \\
            &  $\D_1\pi$   & $5.1 c_1^2 c^2 + 7.6 c^2 s^2 + 0.7 s_1^2 s^2 +$   & $^3D_2 = +0.032 s_1s - 0.087 c_1c - 0.106 c_1s$  \\
            &              & $12.5 c_1^2 cs - 3.8 c_1s_1 cs - 4.6 c_1s_1 s^2 = 1.2$ &    \\
            &  $\D_1'\pi$   & $9.8 c_1^2 c^2 + 6.8 c_1^2 s^2 + 0.9 s_1^2 c^2 - $   & $^3D_2 = +0.035 s_1c + 0.096 c_1 s - 0.115 c_1c$  \\
            &               & $ -16 c_1^2 sc  - 6.0 c_1s_1 c^2 + 5.0 c_1s_1 sc = 7.5$ &        \\
            &  $\D_2\pi$   & $115 c_1^2 + 86 s_1^2 + 195 s_1c_1 = 3.9$   & $^5S_2 = +0.427 c_1 + 0.349 s_1$  \\
            &              &                                     & $^5D_2 = +0.067 c_1 + 0.131 s_1$  \\
            &              &                                     & $^5G_2 = -0.011 c_1 + 0.013 s_1$  \\
            &  $\D_s^* K$  & $45 c_1^2 + 31 s_1^2 + 69 s_1c_1 = 3.3$   & $^3D_2 = -0.174 c_1 - 0.142 s_1$  \\
            &              &                                  & $^3F_2 = 0.086 c_1 - 0.035 s_1$ \\
            &  $\D \rho$  & $100 c_1^2 + 75 s_1^2 + 122 s_1c_1 = 26$   & $^3D_2 = -0.224 c_1 - 0.183 s_1$  \\
            &             &                                    & $^3F_2 = +0.075 c_1 - 0.092 s_1$ \\
            &  $\D \omega$  & $33c_1^2 + 25s_1^2 + 41s_1c_1 = 7.9$   & $^3D_2 = -0.130 c_1 - 0.106 s_1$   \\
            &                &                               & $^3F_2 = +0.042 c_1  - 0.051 s_1$  \\
            &  $\D^* \rho$  & $25 c_1^2 + 50 s_1^2 = 42$   & $^3D_2 = +0.243 s_1$, $^5D_2 = 0.172 c_1$ \\
            &               &                              & $^3F_2 = -0.025 s_1$, $^5F_2 = 0.029 c_1$  \\
            &  $\D^* \omega$  & $7 c_1^2 + 14 s_1^2 = 12$   & $^3D_2 = +0.133 s_1$, $^5D_2 = 0.094 c_1$ \\
            &                &                                  & $^3F_2 = +0.012 s_1$, $^5F_2 = 0.014 c_1$  \\
            &  $\D_{s0} K$ &  $\approx 0$  &  $^1D_2 = -0.003 c_1 + 0.007 s_1$   \\
            & total: & $437 c_1^2 + 389 s_1^2 + 409 s_1c_1 + 9$  &  \\
\hline
\hline
\end{tabular}
\end{table*}

\begin{table*}
\caption{Open-flavor strong decays, continued}
\label{DecTab5}
\vskip 0.3cm
\begin{tabular}{llll}
State &  Mode & $\Gamma_{thy}$\ \  [$\Gamma_{expt}$] (MeV) & Amps. (GeV$^{-1/2}$) \\
\hline
\hline 
$\D''(3.23)$ &  $\D^* \pi$  & 46   & $^3P_0 = +0.103$  \\
             &  $\D_0(2308) \pi$  & 72   & $^1S_0 = +0.154$  \\
             &  $\D_0(2407) \pi$  & 63   & $^1S_0 = +0.156$  \\
             &  $\D^* \eta$  & 1.6   & $^3P_0 = +0.020$  \\
             &  $\D_2 \pi$  & 36   & $^5D_0 = +0.124$  \\  
             &  $\D_s^* K$  & 1.8   & $^3P_0 = +0.022$  \\
             &  $\D \rho$  & 1.4   & $^3P_0 = -0.018$  \\
             &  $\D \omega$  & 0.6   & $^3P_0 = -0.012$  \\
             &  $\D^* \rho$  & 39   & $^3P_0 = +0.105$  \\
             &  $\D^* \omega$  & 13   & $^3P_0 = +0.061$  \\
             &  $\D_{s0} K$  & 47    & $^1S_0 = +0.134$  \\
             &  $\D_0(2308) \eta$  & 11   & $^1S_0 = +0.066$  \\
             &  $\D_s K^*$  & 2.2   & $^3P_0 = -0.026$  \\
             &  $\D_s^* K^*$  & 3   & $^3P_0 = +0.034$  \\
             &  $\D_2 \eta$  & $\approx 0$   & $^5D_0 = -0.004$  \\
               & total:     & 275/266  &     \\
\hline
$\D^{*''}(3.31)$ & $\D\pi$    & 53   & $^1P_1 = -0.100$  \\
             &  $\D^* \pi$  & 59   & $^3P_1 = +0.112$  \\
             &  $\D \eta$  & 5   & $^1P_1 = -0.033$  \\
             &  $\D_s K$  & 14   & $^1P_1 = -0.055$  \\
             &  $\D^* \eta$  & 4   & $^3P_1 = +0.030$  \\
             &  $\D_1 \pi$  & $50 c^2 + 62 s^2 -34 sc = 74$   & $^3S_1 = -0.095 c + 0.135 s$  \\
             &             &                                   & $^3D_1 = -0.096 c - 0.068 s$  \\
             &  $\D_1' \pi$  & $65 c^2 + 53 s^2 + 33 sc = 42 $  & $^3S_1 = +0.134 c + 0.095 s$  \\
             &              &                                  & $^3D_1 = -0.070 c + 0.099 s$  \\
             &  $\D_2 \pi$  & 35   & $^5D_1 = +0.115$  \\
             &  $\D_s^* K$  & 7   & $^3P_1 = +0.041$  \\
             &  $\D \rho$  & 3   & $^3P_1 = +0.024$  \\
             &  $\D \omega$  & 1   & $^3P_1 = +0.013$  \\
             &  $\D^* \rho$  & 12  & $^1P_1 = +0.012$, $^3P_1 = 0$, $^5P_1 = -0.055$, $^5F_1 = 0$   \\
             &  $\D^* \omega$  & 5   & $^1P_1 = +0.007$, $^3P_1 = 0$, $^5P_1 =-0.034$, $^5F_1 = 0$  \\
             &  $\D \eta'$  & 0.5   & $^1P_1 = -0.011$  \\
             &  $\D_s K^*$  & 0.7   & $^3P_1 = -0.013$   \\
             &  $\D_{s1} K$  & $20  c^2 + 34 s^2 - 41 sc = 49$   & $^3S_1 = -0.085 c + 0.120 s$  \\
             &               &                                     & $^3D_1 = -0.039 c - 0.028 s$  \\
             &  $\D_{s1}' K$  & $27 c^2 + 14 s^2 +36 sc = 2$   & $^3S_1 = +0.120 c + 0.085 s$  \\
             &                &                                       & $^3D_1 = -0.017 c + 0.024 s$  \\
             &  $\D_{1} \eta$  & $4.3 c^2 + 7.9 s^2 - 9.8 sc = 11$   & $^3S_1 = -0.041 c + 0.058 s$  \\
             &                &                                     & $^3D_1 = -0.016 c - 0.012 s$  \\
             &  $\D_{1}' \eta$  & $8.3 c^2 + 4.9 s^2 + 9.8 sc = 1$  & $^3S_1 = +0.058 c + 0.041 s$  \\
             &                  &                                     & $^3D_1 = -0.014 c + 0.020 s$  \\
             &  $\D_s^* K^*$  & 7   & $^1P_1 = +0.011$, $^3P_1 = 0$, $^5P_1 = -0.047$, $^5F_1 = 0$  \\
             &  $\D_2 \eta$  & 0.6   & $^1P_1 = +0.017$  \\
             &  $\D_0(2308) \rho$  & 10   & $^3S_1 = +0.066$, $^3D_1 = 0$  \\
             &  $\D_0(2308) \omega$  & 3   & $^3S_1 = +0.038$, $^3D_1 = 0$  \\
             &  $\D_{s0} K^*$  & 0.3   & $^3S_1 = +0.013$, $^3D_1 = 0$  \\
               & total:     & 399/386  &     \\
\hline
\hline
\end{tabular}
\end{table*}

\begin{table*}
\caption{Open-flavor strong decays, continued}
\label{DecTab6}
\vskip 0.3cm
\begin{tabular}{llll}
State &  Mode & $\Gamma_{thy}$\ \  [$\Gamma_{expt}$] (MeV) & Amps. (GeV$^{-1/2}$) \\
\hline
\hline
$\D_{s1}'$ &  $\D^* K$  & $261 c^2 + 131 s^2 - 369 sc = 0.8$ [$<$ 2.3]  & $^3S_1 = 0.789 c + 0.557 s$, $^3D_1 =  -0.024 c + 0.035 s$  \\
\hline
$\D_{s2}$ & $\D K$    & 27  & $^1D_2 = +0.143$ \\
          & $\D^* K$  & 3.1  & $^3D_2 = -0.069$     \\
          & $\D_s \eta$  & 0.2   & $^1D_2 = -0.018$    \\
          & total: &     30 [15(5)]   &        \\
\hline
$\D_{s}'$  & $\D^* K$ & 126  & $^3P_0 = -0.330$ \\
           & $\D_s^*\eta$ &  0.5 & $^3P_0 = +0.043$ \\
           & total: &  127  & \\
\hline
$\D_s^{*'}$ & $\D K$ &  17  & $^1P_1 = +0.090$  \\
            & $\D^* K$ &  81 &  $^3P_1 = -0.236$ \\
            & $\D_s \eta$ & 2.6  & $^1P_1 = -0.042$ \\
            & $\D_s^* \eta$ & 4.1  & $^3P_1 = +0.074$ \\
           & total: &  105  & \\
\hline
\hline
\end{tabular}
\end{table*}

\begin{table*}
\caption{Open-flavor strong decays, continued}
\label{DecTab7}
\vskip 0.3cm
\begin{tabular}{llll}
State &  Mode & $\Gamma_{thy}$\ \  [$\Gamma_{expt}$] (MeV) & Amps. (GeV$^{-1/2}$) \\
\hline
\hline
$\D_s(^3D_1)$ &  $\D K$  & 120   & $^1P_1 = -0.205$   \\
            &  $\D^* K$  & 74   & $^3P_1 = -0.181$  \\
            &  $\D_s \eta$  & 39   & $^1P_1 = +0.129$  \\
            &  $\D_s^*\eta$   & 17   & $^3P_1 = +0.102$  \\
            &  $\D K^*$   & 81   & $^3P_1 = -0.218$  \\
            & total: & 331  &  \\
\hline
$\D_{s2}^*$ &  $\D^* K$  & $155 c_1^2 + 174 s_1^2 - 92 s_1c_1 = 211$  & $^3P_2 = +0.189 c_1 - 0.232 s_1$ \\
            &           &                                    & $^3F_2 = +0.173 c_1 + 0.141 s_1$ \\
            &  $\D_s^*\eta$   & $24 c_1^2 + 35 s_1^2 - 51 s_1c_1 = 55$   & $^3P_2 = -0.112 c_1 + 0.137 s_1$  \\
            &                 &                                  & $^3F_2 = -0.036 c_1 - 0.029 s_1$  \\
            &  $\D K^*$   & $118 c_1^2 + 165 s_1^2 - 230 s_1c_1 = 258$  & $^3P_2 = +0.237 c_1 - 0.290 s_1$  \\ 
            &            &                                      & $^3F_2 = +0.089 c_1 + 0.073 s_1$  \\ 
            &  $\D_0(2308) K$   & $0.04 c_1^2 + 0.78 s_1^2 + 0.34 s_1c_1 = 0.4$  & $^1D_2 = -0.006 c_1 - 0.029 s_1$  \\
            &  $\D_{s0} \eta$   & $0.07 c_1^2 + 0.04 s_1^2 + 0.11 s_1c_1 \approx 0$   & $^1D_2 = +0.010 c_1 + 0.008 s_1$  \\
            &  $\D^* K^*$   & $26 c_1^2 + 13 s_1^2 = 17$   & $^3P_2 = -0.207 c_1$, $^5P_2 = 0.147 s_1$  \\
            &               &                        & $^3F_2 = -0.008 c_1$, $^5F_2 = 0.009 s_1$  \\
            &  $\D_0(2407) K$   & $\approx 0$  & $^1D_2 = -0.003 c_1 - 0.0055 s_1$  \\
            &  $\D_1' K$     &  $\approx 0$  & $^3D_2 = -0.0009 c_1 c + 0.0026 s_1 s -0.0026 s_1 c$  \\
            & total: & $323 c_1^2 + 389 s_1^2 - 373 s_1c_1$ &  \\
\hline
${\D_{s2}^*}'$ &  $\D^* K$  & $174 c_1^2 + 155 s_1^2 + 92 s_1c_1 = 118$   & $^3P_2 = -0.232 c_1 - 0.189 s_1$  \\
             &              &                                  & $^3F_2 = +0.141 c_1 - 0.173 s_1$  \\
            &  $\D_s^*\eta$   & $35 c_1^2 + 24 s_1^2 + 51 s_1c_1 = 4.0$   & $^3P_2 = +0.137 c_1 + 0.112 s_1$  \\
            &                 &                                  & $^3F_2 = -0.029 c_1 + 0.036 s_1$  \\
            &  $\D K^*$   & $165 c_1^2 + 118 s_1^2 + 230 s_1c_1 = 25$   & $^3P_2 = -0.290 c_1 - 0.237 s_1$  \\
            &             &                                     & $^3F_2 = +0.073 c_1 - 0.089 s_1$  \\
            &  $\D_0(2308) K$   & $0.78 c_1^2 + 0.04 s_1^2 - 0.34 s_1c_1 = 0.4$   & $^1D_2 = -0.028 c_1 - 0.0006 s_1$  \\
            &  $\D_{s0} \eta$   & $\approx 0$   & $^1D_2 = +0.008 c_1 - 0.010 s_1$  \\
            &  $\D^* K^*$   & $13 c_1^2 + 26 s_1^2 = 17$   & $^3P_2 = -0.207 s_1$, $^5P_2 = 0.145 c_1$ \\
            &               &                           & $^3F_2 = +0.008 s_1$, $^5F_2 = 0.009 c_1$ \\
            &  $\D_0(2407) K$   & $\approx 0$   & $^1D_2 = -0.005 c_1 + 0.003 s_1$  \\
            &  $\D_1' K$     &  $\approx 0$ & $^3D_2 = -0.0009 s_1 c + 0.003 c_1 s - 0.003 c_1 c$  \\
            & total: & $389 c_1^2 + 323 s_1^2 + 373 s_1c_1$ &  \\
\hline
$\D_s(^3D_3)$ &  $\D K$  & 82   & $^1F_3 = -0.166$  \\
            &  $\D^* K$  & 67   & $^3F_3 = +0.168$  \\
            &  $\D_s \eta$  & 4.5   & $^1F_3 = +0.043$  \\
            &  $\D_s^*\eta$   & 2.2   & $^3F_3 = -0.035$  \\
            &  $\D K^*$   & 14   & $^3F_3 = +0.087$  \\
            &  $\D^* K^*$   & 52   & $^5P_3 = -0.003$, $^1F_3 = 0.294$, $^3F_3 = 0$, $^5F_3 = 0.006$, $^5H_3 = 0$  \\
            & $\D_1' K$     & $\approx 0$ & $^3D_3 = -0.0004 c - 0.0017 s$, $^3G_3 \approx 0$ \\
            & total: & 222 &  \\
\hline
\hline
\end{tabular}
\end{table*}

\begin{table*}
\caption{Open-flavor strong decays, continued}
\label{DecTab8}
\vskip 0.3cm
\begin{tabular}{llll}
State &  Mode & $\Gamma_{thy}$\ \  [$\Gamma_{expt}$] (MeV) & Amps. (GeV$^{-1/2}$) \\
\hline
\hline
$\D_s''(3.23)$ &  $\D^* K$  & 3.5   & $^3P_0 = -0.030$  \\
               &  $\D_s^* \eta$  & 1.9   & $^3P_0 = +0.024$  \\
               &  $\D K^*$  & 39   & $^3P_0 = -0.100$  \\
               &  $\D_0(2308) K$  & 47   & $^1S_0 = +0.134$ \\
               &  $\D_{s0} \eta$  & 13   & $^1S_0 = -0.073$  \\
               &  $\D^* K^*$  & 72   & $^3P_0 = +0.152$  \\
               &  $\D_0(2407) K$  & 35   & $^1S_0 = +0.129$  \\
               &  $\D_2 K$  & 43   & $^5D_0 = -0.151$  \\
               & total:     & 219/207  &     \\
\hline
${\D_s^*}''(3.29)$ &  $\D K$  & 9.6   & $^1P_1 = -0.044$  \\
               &  $\D^* K$  & 0.3   & $^3P_1 = +0.008$  \\
               &  $\D_s \eta$  & 0.4   & $^1P_1 = +0.009$  \\
               &  $\D_s^* \eta$  & 0.4   & $^3P_1 = +0.011$  \\
               &  $\D K^*$  & 16   & $^3P_1 = -0.062$  \\
               &  $\D^* K^*$  & 117   & $^1P_1 = +0.040$, $^3P_1 = 0$, $^5P_1 = -0.180$, $^5F_1 = 0$  \\
               &  $\D_s \eta'$  & 0.5   & $^1P_1 = +0.012$  \\
               &  $\D_1 K$  & $18 c^2 + 31 s^2 - 39 sc = 45$   & $^3S_1 = -0.082 c + 0.117 s$  \\
               &    &                                             & $^3D_1 = +0.035 c + 0.025 s$ \\
               &  $\D_1' K$  & $33 c^2 + 17 s^2 +44 sc = 1.6$   & $^3S_1 = +0.118 c + 0.083 s$  \\
               &            &                                    & $^3D_1 = +0.015 c - 0.021 s$  \\
               &  $\D_2 K$  & 6   & $^1P_1 = -0.054$  \\
               &  $\D_{s1} \eta$ & $3.8 c^2 + 6.9 s^2 - 8.9 sc = 10$ & $^3S_1 = 0.041 c - 0.058 s$ \\
               &                 &                                    & $^3D_1 = -0.015c -0.010s$ \\
               & $\D_{s1}' \eta$ & $3.6 c^2 + 2.4s^2 + 3.5 sc = 1.1$ & $^3S_1 = -0.045c - 0.032 s$ \\
               &                 &                                  & $^3D_1 = -0.015c + 0.022s$ \\
               & total:     & 208 &     \\
\hline
\hline
\end{tabular}
\end{table*}


\begin{thebibliography}{99}
\bibitem{modes}
The decay modes of Fig. \ref{fits2} are as follows. 
[1]  $b_1 \to \omega\pi$,
[2]  $\pi_2 \to f_2 \pi$,
[3]  $K_0 \to K\pi$,
[4]  $\rho \to \pi\pi$,
[5]  $\phi \to K\bar K$,
[6]  $\pi_2 \to \rho\pi$,
[7]  $\pi_2 \to K^*\bar K +cc$,
[8]  $\pi_2\to \omega\rho$,
[9]  $\phi(1680) \to K^*\bar K + cc$,
[10]  $K^* \to K\pi$,
[11]  ${K^*}' \to K\pi$,
[12]  ${K^*}' \to \rho K$,
[13]  ${K^*}' \to K^* \pi$,
[14]  $D^{*+} \to D^0 \pi^+$,
[15]  $\psi(3770) \to D\bar D$,
[16]  $f_2 \to \pi\pi$,
[17]  $f_2 \to K\bar K$,
[18]  $a_2 \to \rho \pi$,
[19]  $a_2 \to \eta \pi$,
[20]  $a_2 \to K\bar K$,
[21]  $f_2' \to K\bar K$,
[22]  $D_{s2} \to DK + D^*K + D_s\eta$,
[23]  $K_2\to K \pi$,
[24]  $K_2 \to K^*\pi$,
[25]  $K_2\to \rho K$,
[26]  $K_2\to \omega K$,
[27]  $\rho_3 \to \pi\pi$,
[28]  $\rho_3 \to \omega\pi$,
[29]  $\rho_3 \to K\bar K$,
[30]  $K_3 \to \rho K$,
[31]  $K_3 \to K^* \pi$,
[32]  $K_3 \to K\pi$.
\end{thebibliography}
\end{document}